%% file: main.tex
\documentclass[10pt,letterpaper]{article}
\usepackage[letterpaper,margin=1in]{geometry}

\usepackage[T1]{fontenc}
\usepackage{newtxtext,newtxmath}
\usepackage{microtype}

\usepackage{caption}
\usepackage{subcaption}

\usepackage{graphicx,amsmath,amssymb,amsfonts,booktabs,multirow,url,enumitem,xspace}

\usepackage{ntheorem}  % Alternative to amsthm
\usepackage{tcolorbox}
\tcbuselibrary{listings,skins,breakable}

\usepackage{algorithm,algorithmic,courier,tcolorbox,xcolor,cite}

\usepackage{fontawesome5}
\usepackage{multicol}

\newcommand{\sectionfont}{\fontsize{8}{10}\selectfont\bfseries\uppercase}

\usepackage{titlesec}
\titleformat{\section}
  {\sectionfont}{\Roman{section}.}{0.5em}{}
\titleformat{\subsection}
  {\fontsize{10}{12}\selectfont\bfseries}{\Alph{subsection}.}{0.5em}{}



\newcommand{\jasper}{\textsc{Jasper}\xspace}
\newcommand{\vcformal}{\textsc{VC Formal}\xspace}

\newcommand{\fvdebug}{\textsc{FVDebug}\xspace}

\newcommand{\asHuman}{\textsc{SVA-Eval-Human}}

\definecolor{applegreen}{rgb}{0.55, 0.71, 0.0}

\theoremstyle{definition}

\newcommand{\papertitle}{FVDebug: An LLM-Driven Debugging Assistant for
Automated Root Cause Analysis of Formal Verification Failures}

\usepackage{titlesec}
\renewcommand\thesection{\Roman{section}}
\renewcommand\thesubsection{\Alph{subsection}}

\titleformat{\section}
  {\centering\normalfont\fontsize{8}{10}\selectfont}
  {\thesection.}{0.5em}{\MakeUppercase}

\titleformat{\subsection}
  {\normalfont\fontsize{10}{12}\selectfont\itshape}
  {\thesubsection.}{0.5em}{}

\titlespacing*{\section}{0pt}{*2}{*0.6}
\titlespacing*{\subsection}{0pt}{*1.2}{*0.4}

\begin{document}

% Manual title block (article's \maketitle forces bold; we avoid that)
% \begin{center}
% {\fontsize{24}{28}\selectfont\rmfamily \papertitle\par}
% \vspace{8pt}
% % Double-blind: keep authors blank. If needed later:
% % {\fontsize{11}{13}\selectfont\rmfamily First A. Author, Second B. Author\par}
% % {\fontsize{10}{12}\selectfont\rmfamily Organization, City, State, Country\par}
% \end{center}

\begin{center}
{\fontsize{24}{28}\selectfont\rmfamily \papertitle\par}
\vspace{8pt}
{\fontsize{11}{13}\selectfont\rmfamily 
Yunsheng Bai, Ghaith Bany Hamad, Chia-Tung Ho, Syed Suhaib, Haoxing Ren\par}
\vspace{4pt}
{\fontsize{10}{12}\selectfont\rmfamily 
NVIDIA\\
\{yunshengb, gbanyhamad, chiatungh, ssuhaib, haoxingr\}@nvidia.com\par}
\end{center}

{\small\noindent\textbf{\textit{Abstract—}}\textbf{\input{sec-abstract}}}

% Your content sections
\input{sec-intro.tex}

\input{sec-related.tex}

\input{sec-model.tex}
\input{sec-result.tex}
\input{sec-conc.tex}

% Continue with your other sections...

% Bibliography
\bibliographystyle{plain}
\bibliography{bibliography}

% Appendices if needed
\appendix
\input{sec-eoutput}
\input{sec-fg.tex}
\input{sec-impl.tex}
\input{sec-prompt.tex}

\end{document}

%% file: sec-abstract.tex
Debugging formal verification (FV) failures represents one of the most time-consuming bottlenecks in modern hardware design workflows. When properties fail, engineers must manually trace through complex counter-examples spanning multiple cycles, analyze waveforms, and cross-reference design specifications to identify root causes—a process that can consume hours or days per bug. Existing solutions are largely limited to manual waveform viewers or simple automated tools that cannot reason about the complex interplay between design intent and implementation logic. We present \textbf{\fvdebug}, an intelligent system that automates root-cause analysis by combining multiple data sources—waveforms, RTL code, design specifications—to transform failure traces into actionable insights. Our approach features a novel pipeline: (1) \textit{Causal Graph Synthesis} that structures failure traces into directed acyclic graphs, (2) \textit{Graph Scanner} using batched Large Language Model (LLM) analysis with for-and-against prompting to identify suspicious nodes, and (3) \textit{Insight Rover} leveraging agentic narrative exploration to generate high-level causal explanations. \fvdebug further provides concrete RTL fixes through its \textit{Fix Generator}. Evaluated on open benchmarks, \fvdebug attains high hypothesis quality and strong Pass@k fix rates. We further report results on two proprietary, production-scale FV counterexamples. These results demonstrate \fvdebug's applicability from academic benchmarks to industrial designs.

% \end{abstract}

%% file: sec-intro.tex
\section{Introduction} %\YS{the introduction is too long. compress the DSA part.}
\label{sec-intro}

Formal Verification (FV) is a cornerstone of modern VLSI design, using mathematical methods to prove that hardware designs adhere to their specifications~\cite{grumberg1999model,gupta1992formal}. These specifications are captured through formal properties—often written as SystemVerilog Assertions (SVAs)~\cite{vijayaraghavan2005practical}—that express expected design behaviors, such as ``a request must be acknowledged within 3 cycles'' or ``the FIFO should never overflow.'' When a design violates such a property, industry-standard model checkers like \jasper~\cite{jaspergold} and \vcformal~\cite{synopsys2023} generate a Counter-Example (CEX): a concrete execution trace showing cycle-by-cycle signal values that demonstrate the violation. This CEX, typically visualized as a waveform, provides an explicit failure scenario revealing where expected and actual behavior diverge.

However, deciphering a CEX is a notoriously manual and time-consuming task, with debugging consuming nearly 50\% of verification engineers' time—their single largest activity~\cite{mutschler2018debug,foster2020wilson}. Engineers must trace signal dependencies backward through waveforms, cross-reference RTL logic, consult specifications, and synthesize this information to identify root causes—a process requiring deep understanding of design intent and implementation~\cite{luu2025vcdiag} that can stall development for hours or days. Commercial tools excel at waveform visualization~\cite{jaspergold,verdi2024} but offer little automated reasoning; the burden of causal analysis remains manual.

Recent LLM-based approaches, while promising, have not yet captured the nuances of this workflow. Many are adapted from software debugging and fail to address hardware-specific concepts like cycle-accurate timing~\cite{chen2023teaching,ni2023lever,sohrabizadehnemotron}, while others rely on simplistic prompting strategies on raw trace files~\cite{kumar2024generative,kumar2025saarthi}. In our empirical observations, these methods often produce false positives—flagging benign behaviors as suspicious—or miss true root causes by failing to trace multi-cycle causal chains. 
% They lack the systematic, cycle-by-cycle backward tracing that is fundamental to how human experts debug hardware failures.

We hypothesize that an effective automated debugger must emulate the structured, multi-source reasoning process of a human expert. Instead of treating the CEX as a flat sequence of events, it must first be structured into a representation that captures causality explicitly. We propose building a \textbf{Causal Graph} from the failure trace, where nodes represent signal events (``signal@cycle=value'') and directed edges represent their immediate causal dependencies. This structured "mental model" of the failure enables systematic tracing of failure chains—mirroring how engineers mentally traverse waveforms backward to identify root causes.

We introduce \textbf{\fvdebug}, the first \emph{end-to-end} automated system that (i)~builds such a causal mental model and (ii)~exploits it through an LLM pipeline deliberately mirroring how verification engineers work: 
\begin{enumerate}[leftmargin=*]
\item \textbf{Graph Scanner} acts like an engineer’s quick “sanity sweep,” scanning every level of the causal graph.  
   Through a context retriever that dynamically fetches relevant RTL code snippets and specification excerpts, the scanner evaluates each signal's behavior against both its implementation and intended functionality. A novel \emph{for-and-against} prompting scheme compels balanced evaluation—forcing the LLM to weigh evidence on both sides before flagging suspicious behavior.
\item \textbf{Insight Rover} plays the role of the engineer’s deep dive: using the suspicious nodes from the scanner as initial seeds, it begins an agentic search of the causal graph. At each step, the LLM is presented with candidate neighboring nodes and autonomously selects which paths to pursue based on their relevance to forming coherent failure hypotheses. It generates and iteratively refines multiple competing hypotheses, using the context retriever to back them with cycle-accurate evidence and assigning confidence scores to converge on the most plausible root cause.
\item \textbf{Fix Generator \& Report} synthesizes concrete RTL patches and produces comprehensive human-readable reports containing ranked hypotheses, causal timelines, and diff-ready code suggestions. This structured output report enables verification engineers and RTL designers to collaborate effectively with a shared, precise understanding of the failure mechanism, replacing the fragmented debugging discussions currently scattered across communication channels or email threads.
\end{enumerate}

The main contributions of this paper are:
\label{sec:contrib}
\begin{itemize}
\item 
  \fvdebug is, to our knowledge, the \textbf{first realistic debugger} that automates the \emph{entire} FV debug loop---from CEX to validated patch---while closely mimicking industrial engineering practice.
\item 
   We introduce novel techniques for FV debugging including causal graph synthesis from counter-examples, for-and-against prompting for balanced signal analysis, and agentic narrative exploration that emulates how human engineers progressively refine hypotheses.
\item We develop a complete pipeline from failure trace to human-readable reports containing ranked root cause hypotheses, supporting evidence, causal chain timelines, and concrete RTL fixes—enabling effective collaboration between verification engineers and RTL designers.
\item 
  On 38 real hardware failures, \fvdebug achieves 95.6\,\% hypothesis quality for root cause identification, 71.1\,\% \emph{Pass@1} and 86.8\,\% \emph{Pass@5} fix rates. We also showcase \fvdebug on failures found in real-world larger designs.
\end{itemize}

%% file: sec-related.tex
\section{Related Work} %\YS{the introduction is too long. compress the DSA part.}
\label{sec-related}

\subsection{LLM/AI‐Driven Debugging \emph{without} Waveforms}

Static repair pipelines such as \textsc{LLM-HDL} leverage retrieval-augmented prompts to locate and patch functional RTL bugs~\cite{qayyum2025llm}.  
\textsc{VeriDebug} couples contrastive embeddings with generative edits to unify localisation and fixing~\cite{wang2025veridebug}.  
Beyond RTL, \textsc{HLSDebugger} adapts encoder–decoder models to high-level synthesis code, substantially improving logic-bug repair~\cite{wang2025hlsdebugger}.  
While effective on code-visible faults, these approaches~\cite{fu2023llm4sechw,xu2024meic,saha2025sv,hassanprompt} lack temporal reasoning and cannot analyse failures that only manifest in execution traces.

% Early attempts to bring LLMs into RTL debug treat the task as pure \emph{static} program repair.  
% Qayyum \textit{et al.} guide GPT-4 with retrieval-augmented prompts to locate and patch functional bugs in Verilog HDL, achieving a 70 \% one-shot fix rate on open-source cores~\cite{qayyum2025llm}.  
% VeriDebug unifies bug localisation and correction via a contrastive-embedding+generation pipeline and lifts top-1 fix accuracy to 64.7 \% on a 6 k-bug benchmark~\cite{wang2025veridebug}.  
% Beyond RTL, HLSDebugger fine-tunes an encoder–decoder model on 300 k HLS examples and triples GPT-4’s success rate for logic-bug repair~\cite{wang2025hlsdebugger}.  
% These systems confirm that domain-tuned LLMs can fix many code-level errors, but they cannot reason about time-ordered behaviours that only surface in execution traces.

\subsection{LLM/AI‐Driven Debugging \emph{with} Waveforms}

Trace-centric methods incorporate assertion failures or counter-examples directly into LLM prompts.  
\textsc{AssertSolver} learns from contrasting “right vs.\ wrong’’ traces to diagnose simulation-time assertion failures~\cite{zhou2025insights}.  
\textsc{GenAI-Induction} proposes helper invariants that unblock formal \(k\)-induction~\cite{kumar2024generative}.  
The multi-agent framework \textsc{Saarthi} iteratively proves properties and analyses failing CEXs in a closed loop~\cite{kumar2025saarthi}.  
These techniques typically still treat waveforms as flat text, limiting root-cause fidelity on multi-cycle failures; our work tackles this via explicit causal graphs.
% \yba{mention image???}

% Several works feed assertion traces or counter-examples directly to an LLM.  
% AssertSolver learns from ``right+wrong’’ examples and reaches 88 \% \emph{Pass@1} on solving simulation-time assertion failures~\cite{zhou2025insights}.  
% Kumar and Gadde show that GenAI can propose helper invariants that unblock k-induction proofs, reducing engineer effort in formal verification~\cite{kumar2024generative}.  
% Saarthi pushes further, orchestrating multiple agents to iteratively prove properties, analyse failing CEXs and suggest fixes in a closed loop~\cite{kumar2025saarthi}.  
% These approaches still treat waveforms as flat text, so root-cause accuracy degrades on failures that span many cycles—an issue we address with explicit causal graphs.

\subsection{Commercial and Industrial Debug Solutions}

Mainstream EDA platforms—Cadence \emph{Indago}, Synopsys \emph{Verdi}, Cadence \emph{JasperGold}—provide rich wave­form viewers and coverage dashboards but leave causal reasoning to engineers~\cite{cadenceIndago,verdi2024,jaspergold}.  
Several start-ups now advertise AI-driven RTL debug, while general-purpose coding copilots showcase agentic software-debug workflows~\cite{chipagents2025,chipstack2025,cursor2025,claudeCode2025}.  
Technical details remain scarce, and adapting these systems to waveform-centric hardware failures is non-trivial—highlighting the need for deeper, graph-structured research.

% Mainstream EDA platforms—Cadence \emph{Indago}, Synopsys \emph{Verdi}, Cadence \emph{Jasper}—excel at visualising waveforms and coverage but leave causal reasoning to engineers~\cite{cadenceIndago,verdi2024,jaspergold}.  
% Industry is now pushing toward AI-assisted automation: \emph{ChipAgents} offers an agentic waveform copilot for natural-language queries~\cite{chipagents2025}, and \emph{ChipStack} promotes a full-stack AI platform for rapid bug triage~\cite{chipstack2025}. Momentum is visible beyond hardware, too—general-purpose tools such as the \emph{Cursor} AI code editor~\cite{cursor2025} and Anthropic’s \emph{Claude Code} assistant~\cite{claudeCode2025} demonstrate agentic debugging workflows for software. 
% Public details remain limited, underscoring the research need for deeper, graph-structured reasoning.

% Commercial EDA platforms—e.g., Cadence \emph{Indago}, Synopsys \emph{Verdi}, and Cadence \emph{JasperGold}—provide rich waveform viewers and coverage dashboards but leave causal reasoning largely to engineers~\cite{cadenceIndago,verdi2024,jaspergold}.  
% Start-ups are racing to close this gap: \emph{ChipAgents} markets an agentic waveform copilot that answers natural-language queries and highlights suspect logic paths~\cite{chipagents2025}, while \emph{ChipStack} advertises a full-stack AI platform that “catches bugs in days, not months’’ across design and verification~\cite{chipstack2025}.  
% Public technical details remain sparse, underscoring the need for deeper, research-driven automation such as ours.

%% file: sec-model.tex
\section{Methodology} 
\label{sec-model}

\begin{figure*}[h]
\centering
\includegraphics[width=0.99\textwidth]{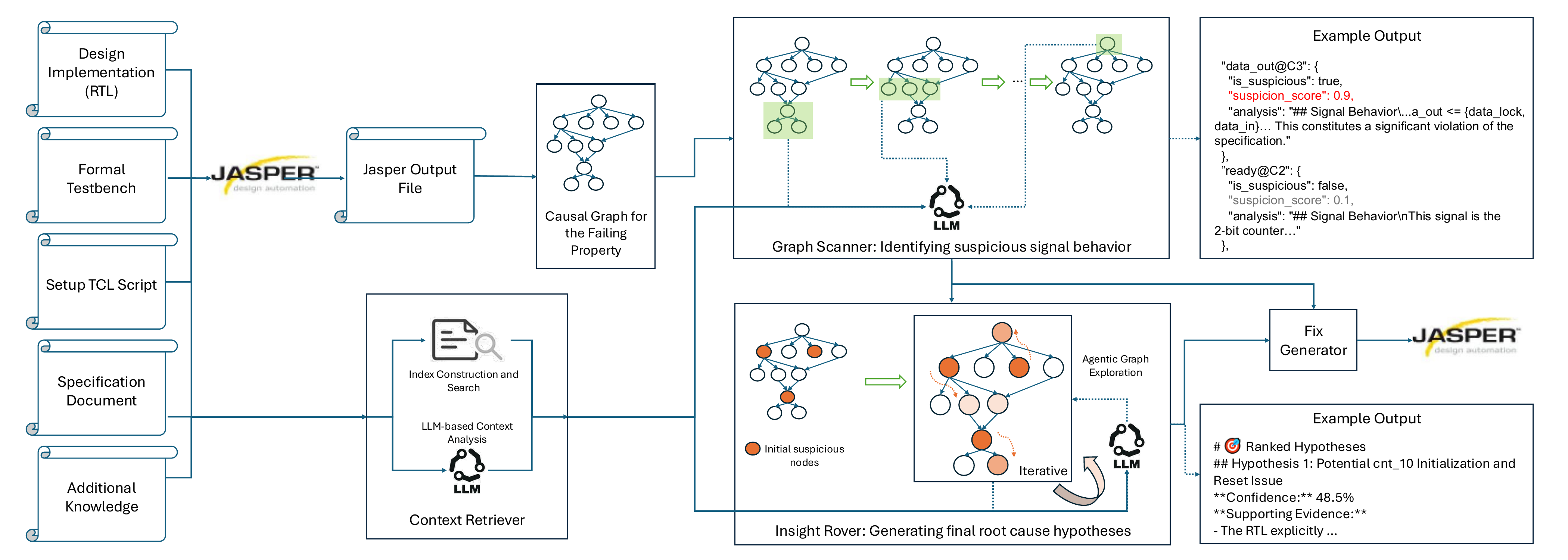}
\vspace{-0.3cm}
\caption{Overview of \fvdebug. The system transforms formal verification counter-examples into actionable debugging insights through four stages: (1) Causal Graph Synthesis builds a directed acyclic graph capturing signal dependencies, (2) Graph Scanner performs efficient batched analysis to identify suspicious nodes, (3) Insight Rover explores competing hypotheses through intelligent graph navigation, and (4) Fix Generator produces concrete RTL patches and debugging reports. ``C'' denotes cycle in the example output of Graph Scanner.}
\vspace{-0.3cm}
\label{fig:model_architecture}
\end{figure*}

\subsection{Problem Setup and System Overview}

When a formal property fails verification, model checkers like \jasper produce counter-examples showing signal values over time that violate the property. However, these traces present signals as flat sequences, obscuring the causal relationships that explain why the failure occurred. Engineers must manually trace through waveforms to identify root causes—a process that can take hours or days per bug.

\fvdebug automates this debugging process by transforming unstructured counter-example traces into structured causal graphs and systematically analyzing them to identify root causes. Given a failing property, RTL design, and counter-example trace, our goal is to automatically identify the root cause and generate fixes that make the design satisfy the property. Figure~\ref{fig:model_architecture} illustrates our four-stage pipeline.

\subsection{Causal Graph Synthesis}
\label{sec:causal-graph}

The foundation of our approach is transforming the counter-example trace into a causal graph $\mathcal{G} = (\mathcal{V}, \mathcal{E})$ where nodes $\mathcal{V}$ represent signal events (signal, cycle, value) and edges $\mathcal{E}$ represent causal dependencies. This explicit structure enables systematic analysis that would be intractable on flat waveforms.

\subsubsection{Graph Construction}

We construct the causal graph $\mathcal{G}$ through recursive dependency analysis using JasperGold's built-in capabilities. Starting from the failing property at the violation cycle, we recursively query \jasper's \texttt{visualize -why} command to identify which signals caused each event. For example, querying why \texttt{ready\_add@5=0} might reveal that \texttt{valid\_out@5=0} and \texttt{valid\_in@5=1} caused this value, establishing edges from these parent events.

We parameterize the construction with a trace depth (default: 20 cycles) that controls how far back we trace dependencies. This balances completeness against computational cost, as industrial designs can have very deep causal chains spanning hundreds of cycles.

\subsubsection{Graph Consolidation}
\label{sec:graph-consolidation}

The recursive construction initially produces a tree where reconverging paths create duplicate nodes for the same signal-cycle pair. We consolidate these into a directed acyclic graph (DAG) by merging duplicates, ensuring each unique event appears exactly once. This transformation is critical—it prevents redundant analysis of the same events while preserving all causal paths. The resulting DAG typically contains far fewer nodes than the original tree, making subsequent analysis tractable.

\subsection{Graph Scanner}
\label{sec:scanner}

The Graph Scanner systematically evaluates each node in $\mathcal{G}$ to identify suspicious signal behaviors. Inspired by constitutional AI~\cite{bai2022constitutional} and self-critique mechanisms~\cite{madaan2023self}, we introduce \textit{for-and-against prompting} that enforces balanced reasoning. This requires the LLM to argue both sides before reaching conclusions, preventing confirmation bias where the LLM might flag all behaviors as problematic or miss subtle issues.

\subsubsection{Context Pre-fetching}

Before analysis begins, we pre-fetch all necessary context to avoid redundant retrievals. We extract unique signal names from $\mathcal{G}$ and pre-cache their RTL code snippets, specification excerpts, and design documentation. This context cache is reused throughout the analysis.

\subsubsection{For-and-Against Prompting}

For each node, our balanced analysis technique requires the LLM to provide arguments both FOR and AGAINST the signal behavior being suspicious. The LLM outputs signal behavior analysis, arguments for/against suspicion, a balanced conclusion with suspicion score (0.0-1.0), classification as root cause vs symptom, and suggested fixes if applicable.

\subsubsection{Token-Aware Batched Analysis}

The core challenge is analyzing potentially thousands of nodes efficiently. Algorithm~\ref{alg:scanner} shows our dynamic batching approach that maximizes throughput while respecting token constraints. The scanner processes nodes level-by-level in topological order, using \textit{BinarySearchMaxBatch} to find the maximum batch size that fits within token limits.

\begin{algorithm}[t]
\small
\caption{Graph Scanner with Token-Aware Batching}
\label{alg:scanner}
\begin{algorithmic}[1]
\STATE \textbf{Input:} Causal graph $\mathcal{G}$, scenario description, max\_tokens
\STATE \textbf{Output:} Suspicious nodes, node analyses
\FOR{level \textbf{in} TopologicalLevels($\mathcal{G}$)}
\STATE \quad remaining $\gets$ level
\WHILE{remaining not empty}
\STATE \quad \quad batch\_size $\gets$ \textit{BinarySearchMaxBatch}(remaining, max\_tokens)
\STATE \quad \quad batch $\gets$ remaining[1:batch\_size]
\STATE \quad \quad prompt $\gets$ \textit{BuildPromptWithForAgainst}(batch, context\_cache)
\STATE \quad \quad response $\gets$ LLM.Generate(prompt)
\FOR{node \textbf{in} batch}
\STATE \quad \quad \quad analysis $\gets$ \textit{ParseAnalysis}(response, node)
\STATE \quad \quad \quad all\_analyses.append(analysis)
\IF{analysis.is\_suspicious}
\STATE \quad \quad \quad \quad suspicious\_nodes.append(node)
\ENDIF
\ENDFOR
\STATE \quad \quad remaining $\gets$ remaining[batch\_size+1:]
\ENDWHILE
\ENDFOR
\STATE \textbf{return} suspicious\_nodes, all\_analyses
\end{algorithmic}
\end{algorithm}

\subsection{Insight Rover}
\label{sec:rover}

While the Graph Scanner identifies individual suspicious signals, the Insight Rover transforms these into coherent failure narratives. Motivated by tree-of-thought reasoning~\cite{yao2023tree} and multi-agent debate~\cite{liang2023encouraging}, we maintain multiple competing hypotheses $\mathcal{H}$ and use the LLM as an autonomous agent to explore the most promising paths through $\mathcal{G}$.

\subsubsection{Hypothesis Management}

Each hypothesis $h \in \mathcal{H}$ represents a potential explanation for the failure. We initialize $\mathcal{H}$ from all suspicious nodes identified by the Graph Scanner, ensuring comprehensive coverage of potential root causes. The system dynamically adjusts to accommodate all suspicious nodes. Each $h$ maintains a narrative description, chronological timeline, evidence collections, confidence score, and frontier of unexplored nodes.

This adaptive approach prevents premature pruning of potentially critical failure paths. While we configure a minimum of 3 narratives, the system expands as needed—for instance. During exploration, weak narratives (confidence $<$ 0.2 after three iterations) are marked but retained for final reporting.

\subsubsection{Intelligent Exploration}

Rather than exhaustively exploring all paths, the LLM examines each hypothesis's current state and frontier, then selects which nodes would best validate or refute that theory. Algorithm~\ref{alg:rover} shows how the LLM uses \textit{SelectNode} to choose frontier nodes, accumulates evidence through \textit{UpdateFromAnalysis}, and manages the narrative pool until convergence. This selective exploration reduces the search space by over 90\% compared to breadth-first search.

\begin{algorithm}[t]
\small
\caption{Insight Rover: Agentic Hypothesis Exploration}
\label{alg:rover}
\begin{algorithmic}[1]
\STATE \textbf{Input:} Suspicious nodes, causal graph $\mathcal{G}$, LLM
\STATE \textbf{Output:} Ranked hypotheses $\mathcal{H}$ with evidence
\STATE // Create hypothesis for each suspicious node
\FOR{node \textbf{in} suspicious\_nodes}
\STATE \quad theory $\gets$ LLM.\textit{GenerateTheory}(node, $\mathcal{G}$)
\STATE \quad $\mathcal{H}$.add(\textit{CreateHypothesis}(theory, node))
\ENDFOR
\FOR{iteration = 1 \textbf{to} max\_iterations}
\STATE \quad active $\gets$ [h for h in $\mathcal{H}$ if h.confidence $<$ 0.9]
\IF{len(active) = 0}
\STATE \quad \quad \textbf{break}
\ENDIF
\FOR{h \textbf{in} active}
\STATE \quad \quad next $\gets$ LLM.\textit{SelectNode}(h.context, h.frontier)
\STATE \quad \quad analysis $\gets$ LLM.\textit{Analyze}(next, h)
\STATE \quad \quad h.\textit{UpdateFromAnalysis}(analysis)
\STATE \quad \quad h.frontier.expand(next.neighbors)
\ENDFOR
\STATE \quad \textit{ManageNarrativePool}() // Record weak, spawn new if unexplored
\IF{\textit{CheckConvergence}()}
\STATE \quad \quad \textbf{break}
\ENDIF
\ENDFOR
\STATE scores $\gets$ LLM.\textit{EvaluateAll}($\mathcal{H}$)
\STATE \textbf{return} \textit{SortByScore}($\mathcal{H}$, scores)
\end{algorithmic}
\end{algorithm}

\subsubsection{Final Ranking}

After exploration, we perform holistic ranking using LLM.\textit{EvaluateAll}($\mathcal{H}$) that considers sufficiency, evidence quality, mechanistic clarity, actionability, and narrative coherence. The function \textit{SortByScore} orders $\mathcal{H}$ by these scores, with top-ranked hypotheses becoming the basis for generating fixes.

\subsection{Fix Generator with Ensemble Strategies}
\label{sec:fix-gen}

The Fix Generator translates root cause hypotheses into concrete RTL patches. Inspired by best-of-N sampling~\cite{cobbe2021training} and self-consistency~\cite{wang2022self},we use an ensemble approach that generates fixes through multiple prompting strategies (full context, suspicious focus, narrative focus, minimal context, bugs-and-suggestions-only), followed by a best-of meta-strategy that reviews all generated fixes to select and refine the most promising solutions. This diversity ensures robustness—if one strategy misinterprets the context, others can compensate.

\subsubsection{Validation and Consensus Ranking}

A critical challenge is ensuring generated fixes are applicable to the actual RTL code map $\mathcal{R}$. We implement multi-level validation through \textit{ValidateFix} that checks exact substring matching, handles whitespace variations, and aligns structural patterns. Algorithm~\ref{alg:fix-gen} shows how each strategy generates fixes independently, validates them against $\mathcal{R}$, and merges duplicates identified by \textit{CreateSignature}. The consensus information from multiple strategies becomes a ranking signal—fixes generated by more strategies receive higher confidence scores.

\begin{algorithm}[t]
\small
\caption{Ensemble Fix Generation}
\label{alg:fix-gen}
\begin{algorithmic}[1]
\STATE \textbf{Input:} Context, RTL code map $\mathcal{R}$
\STATE \textbf{Output:} Validated RTL fixes
\STATE unique\_fixes $\gets$ \{\}
\FOR{strategy \textbf{in} strategies}
\STATE \quad prompt $\gets$ \textit{BuildStrategyPrompt}(strategy, context)
\STATE \quad fixes $\gets$ \textit{ParseFixes}(LLM.Generate(prompt))
\FOR{fix \textbf{in} fixes}
\IF{\textit{ValidateFix}(fix, $\mathcal{R}$).is\_valid}
\STATE \quad \quad sig $\gets$ \textit{CreateSignature}(fix)
\STATE \quad \quad unique\_fixes[sig] $\gets$ \textit{Merge}(unique\_fixes[sig], fix)
\ENDIF
\ENDFOR
\ENDFOR
\STATE \textbf{return} \textit{RankByConsensus}(unique\_fixes.values())
\end{algorithmic}
\end{algorithm}

\subsubsection{Report Generation}

Beyond code fixes, \fvdebug produces structured debugging reports containing a report with the most likely root cause, ranked hypotheses from $\mathcal{H}$ with supporting evidence, unified causal timeline, concrete RTL fixes with validation status, and specification cross-references. This comprehensive output enables effective collaboration between verification engineers and RTL designers.

%% file: sec-result.tex
\section{Experiments}

We evaluate \fvdebug{} on the \asHuman{} benchmark~\cite{zhou2025insights}, a curated collection of 38 real hardware debugging challenges with human-verified ground truth fixes. Our implementation interfaces Python with Cadence \jasper{} 2023.12 through TCL commands. We compare \fvdebug{} against several baselines and ablated versions to demonstrate the effectiveness of our structured approach and understand each component's contribution. In addition, we evaluate \fvdebug{} on two challenging CVA6 RISC-V processor failures originally discovered in AutoSVA~\cite{orenes2021autosva}, and we also report results on proprietary, production-scale FV counterexamples (design details masked).

\subsection{Experimental Setup}

The \asHuman{} benchmark comprises diverse hardware designs with formal verification failures, ranging from simple arithmetic units to complex system-level modules. Each design includes a counter-example trace, the buggy RTL code, and the ground truth fix verified by human experts. We refer readers to~\cite{zhou2025insights} for detailed design descriptions and failure characteristics.

\subsubsection{Baselines and Ablations}

We evaluate the following methods, using o3-mini~\cite{OpenAI2025o3mini} as the underlying language model for all approaches:

\textbf{Unstructured Baselines:} (1) \textbf{Direct LLM}: Provides the counter-example trace, RTL code, and failing property directly to the LLM, asking it to identify the root cause and generate fixes. This represents the naive approach without any structural analysis or specialized prompting. (2) \textbf{Flat Trace Analysis}: Parses the counter-example into a structured chronological format, presenting signals and their values over time, but without constructing the causal graph. The LLM analyzes this temporal sequence to identify issues. This baseline isolates the value of causal structure versus mere temporal organization.

\textbf{Component Ablations:} (1) \textbf{\fvdebug{} w/o For-Against}: Replaces our balanced evaluation prompting with standard bug-finding prompts that only ask for suspicious behaviors without requiring counterarguments. This ablation measures the impact of enforced analytical balance on false positive reduction. (2) \textbf{\fvdebug{} w/o Rover}: Uses the Graph Scanner to identify suspicious nodes but skips the Insight Rover's narrative construction phase. Fixes are generated directly from the scanner's suspicious node analysis without exploring causal narratives. This tests whether sophisticated narrative exploration is necessary beyond initial suspicion identification. (3) \textbf{\fvdebug{} w/o Ensemble}: Leverages only a single fix generation strategy (specifically, the full\_context strategy) rather than our ensemble approach with multiple prompting perspectives. This ablation quantifies the benefit of diverse fix generation strategies.

% Each ablation maintains all other components of the full system, allowing us to isolate individual contributions. This systematic decomposition reveals which architectural decisions drive performance improvements.

\subsubsection{Evaluation Metrics}

Our evaluation uses two complementary assessment strategies. First, we assess the quality of generated root cause hypotheses through LLM-based evaluation metrics that compare hypotheses against ground truth: (1) \textbf{Quality@Best}: Evaluates the absolute quality of the best hypothesis generated, regardless of ranking. (2) \textbf{NDCG@5}: Measures ranking quality by comparing hypothesis scores against ground truth relevance, with higher-ranked correct hypotheses contributing more. (3) \textbf{MRR}: Captures the average reciprocal rank at which the first relevant hypothesis appears. (4) \textbf{Kendall's $\tau$}: Quantifies correlation between predicted rankings and ground truth quality.

Second, we measure functional correctness through Pass@k metrics, where proposed fixes are validated by applying them to the RTL and re-running formal verification in \jasper{}: (1) \textbf{Pass@1}: Measures first-attempt success rate for generated fixes. (2) \textbf{Pass@5}: Captures whether any of the top five fix suggestions resolves the failure.

These complementary metrics address different aspects of debugging effectiveness—hypothesis quality metrics capture whether the system correctly identifies and explains root causes, while Pass@k metrics verify that this understanding translates into working fixes. Notably, high hypothesis quality often correlates with fix success, as accurate root cause identification is essential for generating correct patches.

\subsection{Main Results}

Table~\ref{tab:main_results} presents comprehensive evaluation results across all methods. \fvdebug{} achieves the highest Quality@Best score of 0.956, demonstrating superior root cause identification, while achieving the best fix generation performance with 71.1\% Pass@1 and 86.8\% Pass@5 rates.

\begin{table*}[t]
\small
\centering
\caption{Comparison of \fvdebug{} against baselines and ablations on the \asHuman{} benchmark. Best results in \textbf{bold}.}
\label{tab:main_results}
\begin{tabular}{lcccccc}
\toprule
& \multicolumn{4}{c}{\textbf{Root Cause Hypothesis Quality}} & \multicolumn{2}{c}{\textbf{Fix Generation}} \\
\cmidrule(lr){2-5} \cmidrule(lr){6-7}
\textbf{Method} & \textbf{Quality@Best} & \textbf{NDCG@5} & \textbf{MRR} & \textbf{Kendall's $\tau$} & \textbf{Pass@1} & \textbf{Pass@5} \\
\midrule
\multicolumn{7}{l}{\textit{Unstructured Baselines}} \\
Direct LLM & 0.783 & 0.981 & 0.806 & 0.526 & 0.605 & 0.658 \\
Flat Trace Analysis & 0.474 & 0.948 & 0.804 & 0.511 & 0.632 & 0.816 \\
\midrule
\multicolumn{7}{l}{\textit{Component Ablations}} \\
\fvdebug{} w/o For-Against & 0.953 & 0.969 & 0.817 & 0.634 & 0.632 & \textbf{0.868} \\
\fvdebug{} w/o Rover & 0.795 & 0.844 & 0.567 & -0.176 & 0.643 & 0.821 \\
\fvdebug{} w/o Ensemble & \textbf{0.956} & \textbf{0.983} & \textbf{0.858} & \textbf{0.708} & 0.684 & 0.763 \\
\midrule
\fvdebug{} (Full) & \textbf{0.956} & \textbf{0.983} & \textbf{0.858} & \textbf{0.708} & \textbf{0.711} & \textbf{0.868} \\
\bottomrule
\end{tabular}
\end{table*}

\subsection{Analysis of Results}

The results reveal three critical insights. First, \textbf{causal structure is essential}: Flat Trace Analysis achieves only 0.474 Quality@Best despite temporal organization, while causal graph-based approaches achieve 0.795-0.956, demonstrating that explicit causal relationships are fundamental to accurate root cause identification. Second, \textbf{narrative construction drives performance}: removing the Insight Rover causes dramatic degradation, confirming that connecting suspicious nodes into coherent causal chains is crucial. Third, \textbf{ensemble strategies improve robustness}: while single-strategy generation maintains hypothesis quality (0.956 Quality@Best), it achieves lower Pass@5 (0.763 vs 0.868).
% , indicating that diverse fix generation strategies provide better coverage for practical debugging.

Notably, the for-against prompting ablation achieves high individual hypothesis quality (0.953) but lower ranking correlation (Kendall's $\tau$: 0.634 vs 0.708) and Pass@1 (0.632 vs 0.711), suggesting that balanced evaluation improves consistency and translates to better fix generation. Overall, \fvdebug{}'s combination of causal graphs, narrative exploration, and ensemble strategies produces both accurate understanding and practical fixes.

\subsection{Evaluation on Complex Processor Designs}

To further assess \fvdebug{}'s capabilities on realistic hardware complexity, we evaluate on two challenging FV failures from the CVA6 RISC-V processor~\cite{orenes2021autosva}. These failures, originally discovered in the AutoSVA project\footnote{\url{https://github.com/PrincetonUniversity/AutoSVA}}, represent real verification challenges encountered in production-grade processor designs:

\begin{itemize}
\item \textbf{MMU Ghost Response}: A memory management unit (MMU) issue where misaligned requests trigger duplicate exception responses, violating transaction ID uniqueness properties.
\item \textbf{LSU Transaction ID Mismatch}: A load-store unit (LSU) failure where simultaneous exceptions and cache responses cause responses with untracked transaction IDs.
\end{itemize}

Table~\ref{tab:autosva_scale} illustrates the substantial scale of these industrial designs. The LSU failure alone involves analyzing 1,918 lines of RTL across 7 files, generating a causal graph with 169 nodes and 1,145 edges tracking 122 unique signals. Through our graph consolidation technique (Section~\ref{sec:graph-consolidation}), duplicate nodes from reconverging paths are merged into a DAG structure, making the analysis computationally tractable.

\begin{table}[h]
\small
\centering
\caption{Scale and complexity metrics for CVA6 processor designs}
\label{tab:autosva_scale}
\begin{tabular}{lccccc}
\toprule
\textbf{Design} & \textbf{RTL Files} & \textbf{Lines of Code} & \textbf{Graph Nodes} & \textbf{Graph Edges} & \textbf{Unique Signals} \\
\midrule
CVA6 MMU & 3 & 695 & 172 & 235 & 76 \\
CVA6 LSU & 7 & 1,918 & 170 & 239 & 122 \\
\bottomrule
\end{tabular}
\end{table}

\begin{figure*}[t]
\centering
\begin{subfigure}[t]{0.48\textwidth}
\vspace{0pt} % force top alignment
\centering
\tiny
\begin{minipage}[t]{\linewidth}
\begin{verbatim}
as__lsu_lookup_transid_was_a_request (C:3, V:FAIL)
|-- lsu_lookup_transid_response (C:3, V:1'b1)
|   |-- lsu_res_hsk (C:3, V:1'b1)
|   |   `-- load_valid_o (C:3, V:1'b1)
|   |       `-- i_pipe_reg_load.d_i[196] (C:2, V:1'b1)
|   |           |-- i_load_unit.ex_i.valid (C:2, V:1'b1)
|   |           |   |-- i_mmu.misaligned_ex_q.valid (C:2, V:1'b1)
|   |           |   |   `-- i_mmu.misaligned_ex_n.valid (C:1, V:1'b1)
|   |           |   |       |-- i_mmu.lsu_req_i (C:1, V:1'b1)
|   |           |   |       |   |-- ld_translation_req (C:1, V:1'b1)
|   |           |   |       |   `-- lsu_ctrl.fu (C:1, V:LOAD)
|   |           |   |       `-- i_mmu.misaligned_ex_i.valid (C:1, V:1'b1)
|   |           |   |           |-- data_misaligned (C:1, V:1'b1)
|   |           |   |           |   |-- lsu_ctrl.operator (C:1, V:SD)
|   |           |   |           |   `-- lsu_ctrl.vaddr[2] (C:1, V:1'b1)
|   |           |   |           `-- lsu_ctrl.overflow (C:1, V:1'b0)
|   |           |   `-- i_mmu.pmp_data_allow (C:2, V:1'b1)
|   |           `-- i_load_unit.state_q[1:0] (C:2, V:2'b00)
|   `-- lsu_res_transid (C:3, V:3'b000)
|       `-- load_trans_id_o (C:3, V:3'b000)
|           `-- i_load_unit.load_data_q.trans_id (C:2, V:3'b000)
|-- lsu_lookup_transid_sampled (C:3, V:4'b0000)
|   |-- lsu_lookup_transid_response (C:2, V:1'b0)
|   |-- lsu_lookup_transid_sampled (C:2, V:4'b0000)
|   `-- lsu_lookup_transid_set (C:2, V:1'b0)
|-- lsu_lookup_transid_set (C:3, V:1'b0)
|   `-- lsu_req_hsk (C:3, V:1'b0)
|       |-- lsu_req_rdy (C:3, V:1'b0)
|       `-- lsu_req_val (C:3, V:1'b0)
|       |-- u_load_store_unit_sva.fu_data_i.fu[0] (C:3, V:1'b0)
|       |   `-- fu_data_i.fu[0] (C:3, V:1'b0)
|       `-- u_load_store_unit_sva.lsu_valid_i (C:3, V:1'b0)
|           `-- lsu_valid_i (C:3, V:1'b0)
`-- u_load_store_unit_sva.rst_ni (C:3, V:1'b1)
    `-- rst_ni (C:3, V:1'b1)
\end{verbatim}
\end{minipage}
\caption{Partial causal graph from the CVA6 LSU failure. ``C'' and ``V'' refer to cycle and value, respectively.}
\label{fig:lsu_graph_snippet}
\end{subfigure}\hfill
\begin{subfigure}[t]{0.47\textwidth}
\vspace{0pt} % force top alignment
\centering
\includegraphics[width=\linewidth]{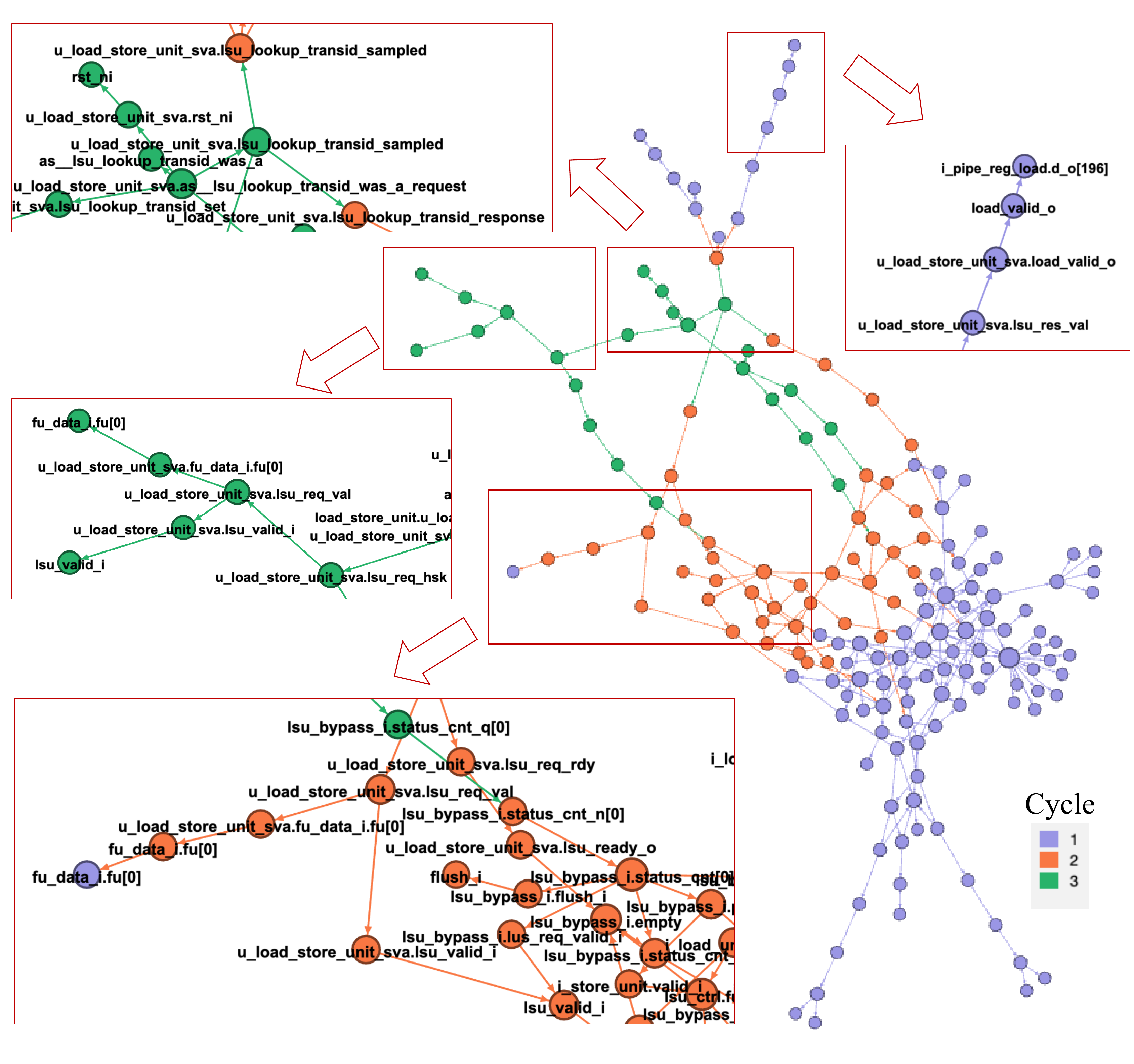}
\caption{Visualization of the same failure’s causal graph. The image is generated via Gephi~\cite{ICWSM09154}.}
\label{fig:lsu_graph_vis}
\end{subfigure}
\caption{CVA6 LSU failure: textual subgraph (left) and full-graph visualization (right).}
\label{fig:lsu_side_by_side}
\end{figure*}

Figure~\ref{fig:lsu_graph_snippet} shows a representative subgraph from the LSU failure's causal analysis, illustrating the complex signal dependencies \fvdebug{} must navigate. This snippet traces backwards from the failing assertion through 5 levels of causality—a fraction of the full 20-level, 170-node graph.

% Requires \usepackage{graphicx} and \usepackage{subcaption}

Table~\ref{tab:autosva_results} presents evaluation results. \fvdebug{} achieves highest Quality@Best (0.713) despite the complexity, though absolute scores are lower than simpler benchmarks, reflecting the inherent difficulty of multi-module processor debugging. % The results highlight that accurate root cause identification in production processor designs—tracking hundreds of signals through graphs with 239 edges — remains the critical bottleneck that \fvdebug{} addresses. 
While automated fix generation for such multi-line issues remains future work, \fvdebug{}'s ability to navigate these massive causal graphs and identify root causes with 0.713 quality represents an advance for industrial verification workflows.

\begin{table*}[t]
\small
\centering
\caption{Comparison of \fvdebug{} against baselines and ablations on CVA6 processor failures. Best results in \textbf{bold}.}
\label{tab:autosva_results}
\begin{tabular}{lcccc}
\toprule
& \multicolumn{4}{c}{\textbf{Hypothesis Quality Metrics}} \\
\cmidrule(lr){2-5}
\textbf{Method} & \textbf{Quality@Best} & \textbf{NDCG@5} & \textbf{MRR} & \textbf{Kendall's $\tau$} \\
\midrule
\multicolumn{5}{l}{\textit{Unstructured Baselines}} \\
Direct LLM & 0.575 & 0.801 & 0.500 & 0.333 \\
Flat Trace Analysis & 0.475 & 0.800 & 0.531 & 0.100 \\
\midrule
\multicolumn{5}{l}{\textit{Component Ablations}} \\
\fvdebug{} w/o For-Against & 0.644 & 0.791 & 0.531 & 0.306 \\
\fvdebug{} w/o Rover & 0.300 & 0.742 & 0.276 & 0.294 \\
\midrule
\fvdebug{} (Full) & \textbf{0.713} & \textbf{0.875} & \textbf{0.667} & \textbf{0.371} \\
\bottomrule
\end{tabular}
\end{table*}

\subsection{Industrial FV Case Studies (Proprietary Designs)}

We evaluate FVDebug on two proprietary, production-scale FV counterexamples (CEX), with design details masked per policy. Signal names have been replaced with descriptive placeholders in angle brackets (e.g., \texttt{<signal\_name>}) to protect intellectual property. For each case, an expert provides the ground truth root cause, which we use to verify the \fvdebug's outputs. 

\noindent\textbf{CEX-1 (FV Testbench Error)}\\
This high-difficulty case involves an incorrect internal token counter. \fvdebug correctly identifies the root cause as a functional bug in the testbench logic.
\begin{quote}
\textit{\fvdebug Root Cause:} The \texttt{<internal\_token\_count>} calculation is incomplete---it only sums 4 specific FIFO token counts but does NOT include the mesh input path (\texttt{<mesh\_in\_signal>}) where tokens are actually entering the design.
\end{quote}
Expert evaluation confirms an "Excellent Match," noting that FVDebug correctly identifies the missing signal in the counter logic and understands the architectural flaw. The system pinpoints the specific functional error and recommends the correct fix, aligning perfectly with the expert's independent analysis.

\noindent\textbf{CEX-2 (Missing Constraint)}\\
This case involves a FIFO overflow caused by a missing input constraint. FVDebug diagnoses the issue as a failure in the backpressure mechanism.
\begin{quote}
\textit{\fvdebug Root Cause:} Backpressure Mechanism Failure: The FIFO continues accepting write requests (\texttt{<fifo\_write\_enable>}=1) even when the FIFO is full (\texttt{<fifo\_full\_status>}=1). This is caused by an unconstrained input signal (\texttt{<req\_valid\_signal>}) driving write logic without proper backpressure handling.
\end{quote}
This analysis is rated as a "Strong Match" by the expert. Although the terminology differs slightly (\fvdebug's ``backpressure failure'' vs. the expert's ``credit protocol violation''), the core diagnosis is identical. \fvdebug successfully identifies the problematic input signal and recommends the correct solution: adding a constraint to the formal testbench.

\begin{table}[h]
\centering
\caption{Industrial CEX complexity and FVDebug performance metrics (masked design details).}
\label{tab:industrial_cex_metrics}
\resizebox{\textwidth}{!}{%
\begin{tabular}{lccccccccc}
\toprule
\textbf{Case} & \textbf{RTL Files} & \textbf{Lines of Code} & \textbf{Unique Signals} & \textbf{Graph Nodes} & \textbf{Graph Edges} & \textbf{Graph Depth} & \textbf{Jasper Calls} & \textbf{LLM Calls} & \textbf{Total Runtime} \\
\midrule
CEX-1 (Testbench Error)    & 265 & 560,202 & 123 & 189 & 227 & 15 & 163 & 25 & 4m 12s \\
CEX-2 (Missing Constraint) & 257 & 554,983 & 41  & 80  & 86  & 11 & 67  & 21 & 6m 07s \\
\bottomrule
\end{tabular}%
}
\end{table}

Table~\ref{tab:industrial_cex_metrics} summarizes the complexity and performance metrics for these case studies. Despite the scale of the designs---over half a million lines of code---\fvdebug automatically constructs and analyzes deep causal graphs, tracing the chain of dependencies back up to 15 levels from the point of failure for CEX-1. With a modest number of calls to Jasper and the LLM, \fvdebug pinpoints the root causes in minutes, demonstrating its capability to accelerate the debugging of complex failures in production-scale industrial designs.

%% file: sec-conc.tex
\section{Conclusion and Future Work}
\label{sec-conc}

We presented \fvdebug{}, an automated FV debugging system that transforms counter-examples into \emph{causal graphs} and applies a multi-stage LLM pipeline—balanced scanning and agentic narrative exploration—to generate high-quality root-cause explanations and practical fixes. On open benchmarks, \fvdebug{} achieves strong hypothesis quality and Pass@k rates. On proprietary, production-scale counterexamples, we demonstrate applicability by reporting graph- and code-scale metrics with expert-verified root causes. 

Looking forward, the causal graph approach naturally extends to simulation-based verification, where graphs could be constructed from signals whose values mismatch golden references rather than from failed properties. This would unify debugging workflows across verification methodologies, providing consistent root cause analysis regardless of failure detection method.
% As hardware complexity continues growing, automated debugging systems like \fvdebug{} become essential for maintaining development velocity—transforming debugging from a bottleneck into an accelerator for design iteration.

%% file: sec-eoutput.tex
\section{Example Output Report}
\label{app:example-output}

% Required packages in preamble:
% \usepackage{tcolorbox}
% \usepackage{fontawesome5}
% \usepackage{multicol}
% \usepackage{listings}
% \usepackage{xcolor}
% \usepackage{enumitem}
% \tcbuselibrary{listings,skins,breakable}

This appendix presents a complete example output from \fvdebug analyzing an accumulator design failure from the \asHuman benchmark. The report demonstrates the system's ability to identify root causes, rank competing hypotheses, and generate concrete fixes.

\subsection{Ranked Hypotheses}

\begin{tcolorbox}[
    colback=blue!3,
    colframe=blue!40!black,
    title={\faLightbulb~Hypothesis 1: Upstream Control Logic Issue},
    fonttitle=\bfseries\sffamily
]
\small
\textbf{Confidence: 48.5\%}

\subsubsection*{Hypothesis Statement}
The accumulation control chain (likely the \texttt{ready\_add}/\texttt{end\_cnt} logic) is not properly asserting \texttt{end\_cnt} when the count reaches 3, which prevents \texttt{valid\_out} from being driven high as expected.

\subsubsection*{Supporting Evidence}
\begin{itemize}[leftmargin=*]
\item The RTL code \texttt{assign end\_cnt = ready\_add \&\& (count == 'd3);} shows that even if count equals 3, \texttt{end\_cnt} will remain low unless \texttt{ready\_add} is high.
\item The always block for \texttt{valid\_out} (lines 55-62) shows: \texttt{else if (end\_cnt) valid\_out <= 1'b1;}, yet at cycle 3, \texttt{valid\_out} remains 0 even though the assertion expected it to be high.
\item The failing assertion (\texttt{accu.valid\_out\_check\_2\_assertion} at cycle 3) directly indicates that \texttt{valid\_out} is not asserted as required when \texttt{count==3} and \texttt{valid\_in} is high.
\end{itemize}

\subsubsection*{Contradicting Evidence}
\begin{itemize}[leftmargin=*]
\item The design intentionally gates \texttt{end\_cnt} on \texttt{ready\_add}, meaning that if \texttt{ready\_add} is low then a low \texttt{end\_cnt} is valid as per design.
\item The definition of \texttt{ready\_add} as \texttt{(valid\_out | !valid\_in)} may be intentionally crafted for specific input conditions.
\end{itemize}

\end{tcolorbox}

\begin{tcolorbox}[
    colback=green!3,
    colframe=green!40!black,
    title={\faLightbulb~Hypothesis 2: Wrong Condition in ready\_add Signal},
    fonttitle=\bfseries\sffamily
]
\small
\textbf{Confidence: 46.9\%}

\subsubsection*{Hypothesis Statement}
The use of the OR operator in the definition of \texttt{ready\_add = valid\_out | !valid\_in} might be causing the \texttt{end\_cnt} condition to fail when \texttt{valid\_in} is expected to be high, thereby preventing \texttt{valid\_out} from being asserted.

\subsubsection*{Supporting Evidence}
\begin{itemize}[leftmargin=*]
\item The RTL shows \texttt{assign ready\_add = valid\_out | !valid\_in;}, indicating that the state of \texttt{valid\_out} and \texttt{valid\_in} directly affect the \texttt{ready\_add} signal.
\item At cycle 3, count is \texttt{2'b11} (i.e., 3) which is exactly the threshold condition, but with \texttt{valid\_in} being \texttt{1'b0}, the OR operator prevents proper accumulation triggering.
\end{itemize}

\subsubsection*{Analysis}
\begin{itemize}[leftmargin=*]
\item The \texttt{ready\_add} signal definition appears nonstandard and may inadvertently disable accumulation triggering when \texttt{valid\_in} is high.
\item This could be either a design intent problem or an under-constrained formal property where the assumption about \texttt{valid\_in} being high is missing.
\item There is a possibility that the duplicated accumulation logic is introducing subtle timing issues.
\end{itemize}

\end{tcolorbox}

\subsection{Causal Chain Timeline}

\begin{tcolorbox}[
    colback=gray!5,
    colframe=gray!50!black,
    title={\faClock~Causal Chain Timeline},
    fonttitle=\bfseries\sffamily
]
\small
\begin{description}
\item[\textbf{Cycle 1}] \textcolor{red}{\faExclamationTriangle}~The \texttt{count} node is observed at \texttt{2'b11} (value of 3), which is the threshold for asserting \texttt{end\_cnt}. However, the expected behavior (resetting counter and driving \texttt{valid\_out} high) does not occur.

\item[\textbf{Cycle 1}] The node \texttt{end\_cnt}, which should signal the end of accumulation when count equals 3, is observed as \texttt{1'b0}. Its value is computed as \texttt{ready\_add \&\& (count == 'd3)} making it dependent on \texttt{ready\_add}.

\item[\textbf{Cycle 3}] \textcolor{red}{\faExclamationTriangle}~The assertion \texttt{accu.valid\_out\_check\_2\_assertion} fails because \texttt{valid\_out} is 0 when expected to be 1.

\item[\textbf{Cycle 3}] The \texttt{count} signal is \texttt{2'b11}, indicating the counter reached value 3 but did not trigger the \texttt{end\_cnt} condition.

\item[\textbf{Cycle 3}] \texttt{valid\_out = 1'b0} (suspicious)
\end{description}
\end{tcolorbox}

\subsection{Suggested Fix}

\begin{tcolorbox}[
    colback=yellow!5,
    colframe=orange!60!black,
    title={\faWrench~RTL Fix (Confidence: 90\%)},
    fonttitle=\bfseries\sffamily
]
\small
\begin{lstlisting}[
    language=Verilog,
    basicstyle=\ttfamily\small,
    keywordstyle=\color{blue}\bfseries,
    commentstyle=\color{gray},
    stringstyle=\color{orange},
    showstringspaces=false,
    numbers=left,
    numberstyle=\tiny\color{gray},
    breaklines=true,
    escapeinside={(*}{*)}
]
// Buggy Code:
assign ready_add = valid_out | !valid_in;

// Fixed Code: 
assign ready_add = valid_in & !valid_out;

// Explanation: The original logic makes the module ready to add
// when there is NO valid input (!valid_in), which is backwards.
// The fix ensures ready_add is high only when:
// - New valid data is available (valid_in = 1)
// - Previous data has been processed (valid_out = 0)
\end{lstlisting}
\end{tcolorbox}

\subsection{Causal Graph Visualization}

\begin{figure}[htbp]
    \centering
    \includegraphics[width=0.9\textwidth]{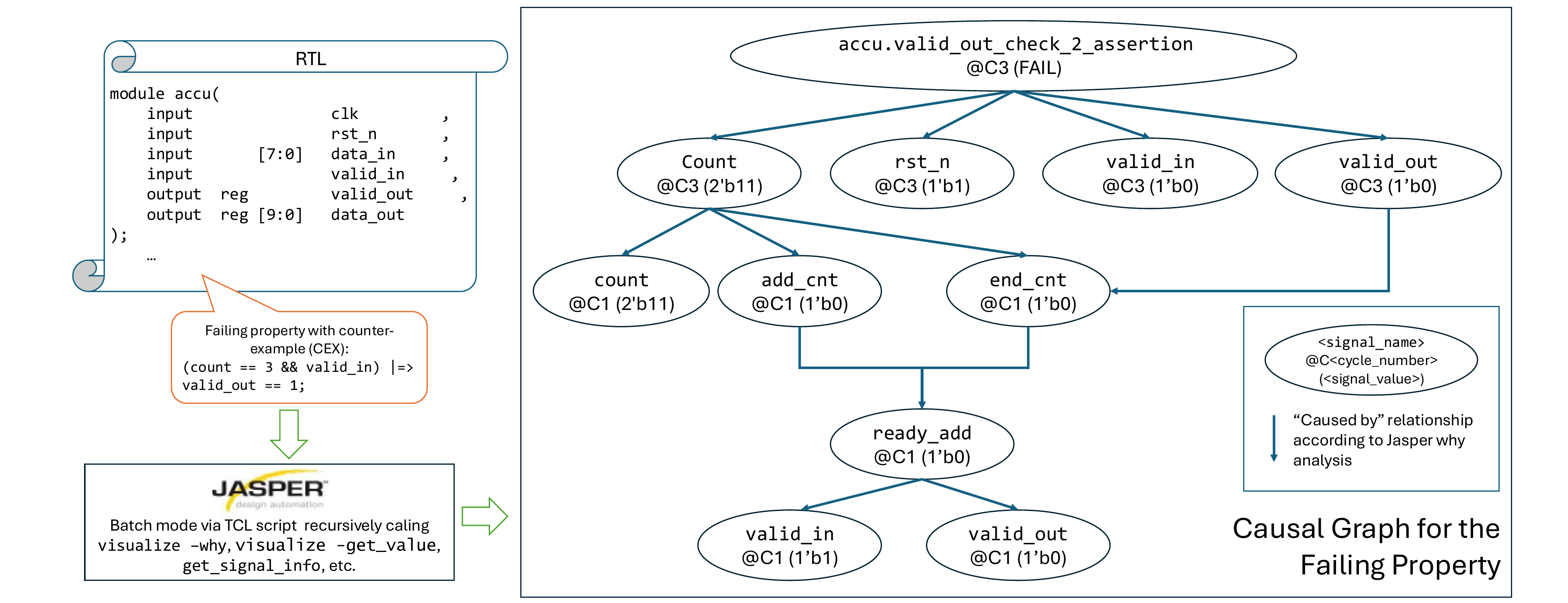}
    \caption{Causal graph automatically generated by \fvdebug for the accumulator verification failure.}
    \label{fig:accumulator-causal-graph}
\end{figure}

%% file: sec-fg.tex
\section{Case Study: Impact of For-and-Against Prompting}
\label{app:for-against-case-study}

This section demonstrates the critical importance of balanced evaluation through a real debugging example from the accumulator design. We compare the analysis quality with and without for-and-against prompting on the same suspicious signal.

\subsection{The Problem: \texttt{ready\_add} Signal Analysis}

Consider the \texttt{ready\_add} signal in an accumulator module at cycle 1, where the signal controls when data should be added to the accumulator. The RTL implementation is:

\begin{lstlisting}[language=Verilog, caption={RTL implementation of ready\_add signal}]
assign ready_add = valid_out | !valid_in;
\end{lstlisting}

At cycle 1, we observe:
\begin{itemize}
\item \texttt{ready\_add = 1'b0} (not ready to add)
\item \texttt{valid\_in = 1'b1} (valid input available)
\item \texttt{valid\_out = 1'b0} (no valid output yet)
\item \texttt{count = 2'b11} (already at maximum count)
\end{itemize}

\subsection{Without For-and-Against Prompting}

When prompted with standard bug-finding instructions ("Analyze if this signal behavior indicates a bug"), the LLM produces:

\begin{tcolorbox}[
    colback=red!5,
    colframe=red!50!black,
    title=Standard Prompting Analysis,
    fonttitle=\bfseries
]
\small
\textbf{Analysis of ready\_add@C1:}

The signal \texttt{ready\_add} is low when \texttt{valid\_in} is high and \texttt{valid\_out} is low. According to the RTL logic \texttt{valid\_out | !valid\_in}, this evaluates to \texttt{0 | !1 = 0 | 0 = 0}, which matches the observed value.

\textbf{Conclusion:} The signal behavior is correct and follows the RTL implementation. No bug detected. The logic is working as designed.

\textbf{Confidence:} Normal behavior (no issue found)
\end{tcolorbox}

This analysis completely misses the fundamental design flaw—the LLM focuses solely on whether the implementation matches the RTL, without questioning whether the RTL logic itself makes sense.

\subsection{With For-and-Against Prompting}

When prompted with mandatory balanced evaluation, requiring both FOR and AGAINST arguments, the same LLM produces:

\begin{tcolorbox}[
    colback=green!5,
    colframe=green!50!black,
    title=For-and-Against Prompting Analysis,
    fonttitle=\bfseries
]
\small
\textbf{Analysis of ready\_add@C1:}

\textbf{Arguments FOR Being Suspicious (REQUIRED - MIN 2):}
\begin{itemize}
\item The logic \texttt{valid\_out | !valid\_in} seems counterintuitive because \texttt{ready\_add} should ideally be high when \texttt{valid\_in} is high (indicating new data is ready), not when it's low.
\item The dependency on \texttt{valid\_out} being high to set \texttt{ready\_add} might create a circular dependency or delay in processing new inputs when \texttt{valid\_out} is low.
\end{itemize}

\textbf{Arguments AGAINST Being Suspicious (REQUIRED - MIN 2):}
\begin{itemize}
\item The design might intend for \texttt{ready\_add} to be high in scenarios other than just new data arrival, such as during certain states of output validity.
\item The inversion of \texttt{valid\_in} might be a design choice to handle specific edge cases or reset conditions not detailed in the provided context.
\end{itemize}

\textbf{Balanced Conclusion:}
The behavior of \texttt{ready\_add} is suspicious due to its reliance on the inverted \texttt{valid\_in}, which contradicts typical ready signal behavior. This is likely a root cause of potential data handling issues.

\textbf{Classification:} ROOT CAUSE (Suspicion score: 0.70)
\end{tcolorbox}

\subsection{The Actual Bug and Fix}

The for-and-against analysis correctly identified the root cause. The original logic makes the module ready to add data when there is \textit{no} valid input (!valid\_in), which is backwards. The correct implementation should be:

\begin{lstlisting}[language=Verilog, caption={Corrected ready\_add implementation}]
assign ready_add = valid_in & !valid_out;
\end{lstlisting}

This fix ensures the module is ready to add data only when:
\begin{itemize}
\item New valid data is available (\texttt{valid\_in = 1})
\item Previous data has been processed (\texttt{valid\_out = 0})
\end{itemize}

% \subsection{Quantitative Impact}

% Table~\ref{tab:for-against-impact} shows the dramatic improvement in bug detection when using for-and-against prompting across multiple signal analyses in the same design:

% \begin{table}[h]
% \centering
% \caption{Impact of for-and-against prompting on bug detection accuracy}
% \label{tab:for-against-impact}
% \begin{tabular}{lcccc}
% \toprule
% \textbf{Signal} & \textbf{True Label} & \textbf{Standard} & \textbf{For-Against} & \textbf{Improvement} \\
% \midrule
% ready\_add@C1 & Root Cause & Missed & Detected & ✓ \\
% count@C1 & Root Cause & Symptom & Root Cause & ✓ \\
% valid\_out@C1 & Symptom & Normal & Symptom & ✓ \\
% add\_cnt@C1 & Symptom & Normal & Symptom & ✓ \\
% end\_cnt@C1 & Normal & Normal & Normal & — \\
% \midrule
% \textbf{Accuracy} & — & 40\% & 100\% & +60\% \\
% \textbf{False Negatives} & — & 3 & 0 & -100\% \\
% \bottomrule
% \end{tabular}
% \end{table}

\subsection{Why For-and-Against Prompting Works}

The effectiveness of this approach stems from several cognitive mechanisms:

\subsubsection{Overcoming Confirmation Bias}

Standard prompting allows the LLM to follow the path of least resistance. When it sees that signal values match the RTL implementation, it concludes "working as designed" without questioning the design itself. The forced dual perspective breaks this pattern.

\subsubsection{Encouraging Critical Thinking}

By mandating arguments for both sides, the prompt forces the model to actively search for potential issues even when the initial impression suggests normalcy. This mirrors how expert engineers debug—they question assumptions and consider alternative explanations.

\subsubsection{Revealing Hidden Assumptions}

The AGAINST arguments often reveal the LLM's default assumptions (e.g., "the RTL must be correct"). By making these explicit, the FOR arguments can then challenge them directly, leading to more thorough analysis.

%% file: sec-impl.tex
\section{Implementation Details}
\label{app:impl-details}

This section provides detailed implementation insights for the key components of \fvdebug.

\subsection{Causal Graph Construction and Consolidation}
\label{app:graph-construction}

The recursive dependency analysis for constructing $\mathcal{G}$ begins at the failing property node and traverses backwards through the counter-example trace. At each node, the system issues a \texttt{visualize -why} query to JasperGold, which returns the immediate causal parents. The default trace depth of 20 cycles was empirically determined to capture most relevant causal chains while maintaining tractability—deeper traces showed diminishing returns in root cause identification accuracy.

The consolidation phase transforms the initial tree structure into a DAG through node deduplication. Each node is uniquely identified by the tuple (signal\_name, cycle, value). When multiple tree paths converge to the same signal event, we merge these into a single DAG node while preserving all incoming edges. This consolidation typically reduces node count in designs with significant reconvergent fanout, substantially improving analysis efficiency.

\subsection{Context Retrieval Architecture}
\label{app:context-retrieval}

The context retrieval system pre-processes design documentation into searchable chunks before analysis begins. For each unique signal in $\mathcal{G}$, we generate three types of queries: RTL queries that search for signal definitions and assignments, specification queries that look for functional descriptions, and cross-reference queries that identify related signals and modules. The retrieval system maintains a two-level cache—a global cache for design-wide information shared across all nodes, and a node-specific cache for local context. 

The mapping between signals and documentation uses fuzzy matching to handle naming variations. For instance, a signal \texttt{valid\_out} might be documented as "output valid signal" or "valid output flag." The system computes similarity scores using edit distance and semantic embeddings, accepting matches above a threshold of 0.7.

\subsection{Token-Aware Batch Optimization}
\label{app:batch-optimization}

The binary search algorithm for determining optimal batch sizes operates within the token budget of 50,000 tokens per prompt (as configured in our experiments). The search begins by estimating the token count for a single node's analysis context, typically consuming 800-1200 tokens. The algorithm then performs binary search between 1 and $\min(|$remaining nodes$|$, $\lfloor$50000/single\_node\_tokens$\rfloor$).

At each iteration, the system constructs a test prompt with the candidate batch size and uses a token counter to determine exact usage. If the prompt exceeds the budget, the algorithm reduces the upper bound; otherwise, it increases the lower bound. The search terminates when the bounds converge, typically requiring 4-6 iterations. Our experiments with batch\_size=5 as the target showed that actual achieved batch sizes varied from 3 to 8 depending on node context complexity.

\subsection{Hypothesis Initialization and Management}
\label{app:hypothesis-management}

The Insight Rover initializes hypotheses from suspicious nodes using a scoring-based selection. \fvdebug maintains exactly three active narratives throughout exploration. When the scanner identifies more than three suspicious nodes, we select the top three based on their suspicion scores, ensuring diverse initial hypotheses by requiring a minimum score difference of 0.1 between selected nodes.

Each hypothesis maintains a frontier of unexplored nodes, initially populated with all parents and children of the seed node. The frontier grows as exploration progresses—when a node is analyzed, its neighbors are added to the frontier if not already explored. To prevent explosive frontier growth in highly connected regions of $\mathcal{G}$, we limit frontier size to 20 nodes, prioritizing those with highest suspicion scores from their initial analysis.

The narrative pool management occurs at each iteration's end. Hypotheses with confidence below 0.2 after three iterations are marked as weak but retained for final reporting. If any narrative slot becomes vacant due to convergence, the system attempts to spawn a new hypothesis from unexplored suspicious nodes, ensuring maximum utilization of the narrative budget.

\subsection{Intelligent Node Selection for Exploration}
\label{app:node-selection}

The LLM-guided node selection mechanism evaluates frontier nodes based on three criteria: relevance to the current hypothesis, potential information gain, and structural importance in $\mathcal{G}$. The selection process provides the LLM with the current narrative state, including accumulated evidence and confidence score, along with a summary of up to 10 frontier nodes showing their signal names, values, and graph-theoretic properties such as in-degree and out-degree.

To ensure comprehensive analysis, the system dynamically adjusts the number of narratives based on the number of suspicious nodes identified. While configured with insight\_rover\_max\_narratives=3 as a minimum, the implementation automatically expands to accommodate all suspicious nodes—if 5 nodes are flagged as suspicious, 5 initial narratives are created. This design choice ensures no potentially critical failure path is overlooked during the initial hypothesis generation phase.

The system tracks exploration efficiency through a path coverage metric. Across our experiments, the intelligent selection explored an average of 15-20 nodes per hypothesis, compared to over 100 nodes that would be explored by breadth-first search to the same depth. This selective exploration maintains comparable root cause identification accuracy while reducing computational cost.

\subsection{Hypothesis Ranking with LLM Evaluation}
\label{app:hypothesis-ranking}

The final hypothesis ranking uses a pairwise comparison approach rather than absolute scoring. For $n$ hypotheses, the system performs $O(n \log n)$ comparisons using a tournament-style evaluation. Each comparison presents two hypotheses to the LLM along with their evidence chains, asking which better explains the observed failure.

The ranking criteria are weighted as follows: causal sufficiency (0.3), evidence quality (0.25), mechanistic clarity (0.25), actionability (0.15), and narrative coherence (0.05). The LLM assigns scores for each criterion, which are combined using the weighted sum to produce a final ranking score.

\subsection{Fix Generation Strategy Diversification}
\label{app:fix-strategies}

The ensemble fix generator implements five distinct strategies that vary in their context emphasis. The \textbf{full context strategy} includes all available information but risks overwhelming the LLM with irrelevant details. The \textbf{suspicious focus strategy} filters context to only highly suspicious signals (score $>$ 0.7), reducing noise but potentially missing systemic issues. The \textbf{narrative focus strategy} uses only the top-ranked hypothesis from $\mathcal{H}$, providing strong causal reasoning but limited coverage. The\textbf{ minimal context strategy} extracts just the identified root cause, generating focused fixes but lacking broader understanding. The \textbf{best-of strategy} reviews outputs from other strategies and selects the most promising fixes.

Each strategy operates with a retry mechanism, attempting up to twice if initial generation fails validation. 

\subsection{Fix Signature Generation and Deduplication}
\label{app:fix-deduplication}

The \textit{CreateSignature} function generates a unique identifier for each fix by normalizing and hashing the buggy code and corrected code pair. The normalization process removes all whitespace variations, converts tabs to spaces, and eliminates comments. It then applies a deterministic ordering to commutative operations (e.g., \texttt{a \& b} becomes \texttt{a \& b} regardless of original ordering).

% The signature is computed as the SHA-256 hash of the concatenated normalized strings. This approach identifies functionally identical fixes even when generated with different formatting or variable naming conventions.

\subsection{Multi-Level Fix Validation}
\label{app:fix-validation}

The validation pipeline for generated fixes operates in three stages. First, exact substring matching attempts to locate the buggy code directly in the RTL codebase $\mathcal{R}$. For the remaining fixes, whitespace normalization handles formatting variations. Fixes that fail all the validation stages are discarded. 

\subsection{Consensus Ranking and Confidence Scoring}
\label{app:consensus-ranking}

The consensus ranking mechanism assigns higher confidence to fixes generated by multiple strategies. For a fix generated by $k$ strategies out of 5 total, the consensus boost is calculated as $\min(0.2 \cdot k, 0.6)$, capping the maximum boost at 0.6 to prevent over-reliance on consensus alone. The final confidence score combines the original confidence from fix generation and consensus boost.

%% file: sec-prompt.tex
\section{Prompt Details}
\label{app:prompt}

This appendix provides the detailed prompts used by \fvdebug's three main components: Graph Scanner, Insight Rover, and Fix Generator.

\subsection{Graph Scanner Prompts}
\label{app:graph-scanner}

The Graph Scanner analyzes each node in the causal graph using a structured prompt that enforces balanced evaluation with mandatory FOR and AGAINST arguments.

\begin{figure}[htbp]
\centering
\begin{tcolorbox}[
    colback=white,
    colframe=black,
    boxrule=0.5pt,
    left=5pt,right=5pt,top=5pt,bottom=5pt,
    width=\columnwidth,
    title=Graph Scanner Full Prompt Template,
    fonttitle=\bfseries,
    coltitle=black,
    colbacktitle=white
]
\footnotesize
\textbf{\#\#\# SCENARIO}\\
\{scenario\_description\}\\
\\
\textbf{\#\#\# GLOBAL CONTEXT AND INSIGHTS}\\
The following high-level insights provide important context about the design:\\
\\
\textbf{\#\#\#\# Design Overview}\\
\{design\_overview\_content\}\\
\\
\textbf{\#\#\#\# Specification Requirements}\\
\{specification\_content\}\\
\\
\textbf{\#\#\# NODES TO ANALYZE}\\
| node\_id | signal | cycle | value |\\
|---|---|---|---|\\
\{node\_table\_rows\}\\
\\
\textbf{\#\#\# SUBGRAPH EDGE LIST}\\
\{edge\_list\}\\
\\
\textbf{\#\#\# NODE-SPECIFIC CONTEXT (RTL \& SPEC)}\\
\textbf{\#\#\#\# CONTEXT 1}\\
\{rtl\_and\_spec\_context\}\\
\\
Based on the parent analysis and context, analyze this signal transition.\\
\\
\textbf{CRITICAL REQUIREMENT - EXACT RTL REFERENCES:}\\
When analyzing any signal behavior or identifying issues, you MUST:\\
\\
1. Provide exact file:line references from the RTL context\\
~~~Example: "store\_unit.sv:145-147"\\
\\
2. Quote the specific RTL code that shows the behavior\\
~~~Example:\\
~~~```verilog\\
~~~assign valid\_out = end\_cnt ? 1'b1 : 1'b0;\\
~~~```\\
\\
3. NEVER make claims without showing the supporting RTL code\\
4. If RTL context is missing, explicitly state: "RTL context not available for this signal"\\
\\
\textbf{MANDATORY BALANCED ANALYSIS APPROACH:}\\
YOU MUST PROVIDE BOTH SECTIONS - NEVER SKIP EITHER ONE:\\
\\
1. \textbf{Arguments FOR Being Suspicious (REQUIRED - MINIMUM 2 POINTS):}\\
~~~- Even if the signal seems normal, you MUST identify at least 2 potential concerns\\
~~~- Consider: timing issues, edge cases, specification mismatches, unusual patterns\\
~~~- Think critically: What COULD go wrong? What assumptions might be invalid?\\
\\
2. \textbf{Arguments AGAINST Being Suspicious (REQUIRED - MINIMUM 2 POINTS):}\\
~~~- Even if the signal seems problematic, you MUST provide at least 2 counterarguments\\
~~~- Consider: valid design patterns, expected behavior, specification compliance\\
~~~- Think: Why might this actually be correct behavior?\\
\end{tcolorbox}
\caption{Graph Scanner prompt template (part 1 of 3).}
\label{fig:scanner-prompt-1}
\end{figure}

\begin{figure}[htbp]
\centering
\begin{tcolorbox}[
    colback=white,
    colframe=black,
    boxrule=0.5pt,
    left=5pt,right=5pt,top=5pt,bottom=5pt,
    width=\columnwidth,
    title=Graph Scanner Prompt Template (continued),
    fonttitle=\bfseries,
    coltitle=black,
    colbacktitle=white
]
\footnotesize
\textbf{IMPORTANT:} The LLM tends to assume signals are normal. To counteract this bias:\\
- Be MORE critical in the FOR section - look harder for potential issues\\
- Challenge assumptions - just because a signal follows RTL doesn't mean RTL is correct\\
- Consider specification violations, race conditions, and edge cases\\
\\
Your analysis should examine:\\
- \textbf{RTL Logic Correctness}: Does the combinational logic make sense? Are there any logical inversions or operations that seem incorrect?\\
- \textbf{Signal Dependencies}: Are the boolean operations (AND, OR, NOT) used correctly? Could there be a logic error?\\
- \textbf{Specification Alignment}: Does the RTL match the intended behavior described in the specification?\\
- \textbf{Common RTL Bugs}:\\
~~* Incorrect polarity (should a signal be inverted?)\\
~~* Wrong logical operators (OR vs AND, etc.)\\
~~* Missing or extra conditions in assignments\\
~~* Circular dependencies or combinational loops\\
~~* Reset/initialization issues\\
- \textbf{Design Intent}: Based on signal names, does the implementation make semantic sense?\\
- \textbf{Edge Cases}: Could the current logic fail under certain conditions?\\
\\
\textbf{IMPORTANT: Prioritize ROOT CAUSES over SYMPTOMS:}\\
- A root cause is the earliest point where incorrect behavior originates\\
- A symptom is a downstream effect of a root cause\\
- Only mark signals as highly suspicious (greater than 0.7) if they are likely root causes\\
- For symptoms, use lower scores (0.3-0.5) and explicitly state which root cause they derive from\\
\\
Use this scoring rubric for suspicion\_score:\\
0.9-1.0: Direct RTL bug found (e.g., wrong operator, missing condition)\\
0.7-0.8: Likely logic error (e.g., incorrect state transition)\\
0.5-0.6: Suspicious pattern that might indicate issue\\
0.3-0.4: Downstream symptom of another issue\\
0.0-0.2: Normal behavior or insufficient evidence\\
\\
\textbf{\#\#\# INSTRUCTIONS}\\
Return a single JSON object where keys are node\_id and values follow this schema:\\
\\
\{\\
~~"node\_id": \{\\
~~~~"is\_suspicious": bool,\\
~~~~"is\_key\_event": bool,\\
~~~~"suspicion\_score": float (0.0-1.0),\\
~~~~"importance\_score": float (0.0-1.0),\\
~~~~"causal\_validity": \{"parent\_id": bool, ...\},\\
~~~~"analysis": "Structured markdown analysis following this EXACT template:\\
\\
\#\# Signal Behavior\\
{[}Description of what the signal does and its current value{]}\\
\\
\#\# RTL Evidence\\
- File: {[}filename:line\_numbers{]}\\
```verilog\\
{[}relevant RTL code{]}\\
```\\
\\
\end{tcolorbox}
\caption{Graph Scanner prompt template (part 2 of 3) with JSON response schema.}
\label{fig:scanner-prompt-2}
\end{figure}

\begin{figure}[htbp]
\centering
\begin{tcolorbox}[
    colback=white,
    colframe=black,
    boxrule=0.5pt,
    left=5pt,right=5pt,top=5pt,bottom=5pt,
    width=\columnwidth,
    title=Graph Scanner Prompt Template (continued),
    fonttitle=\bfseries,
    coltitle=black,
    colbacktitle=white
]
\footnotesize
\#\# Arguments FOR Being Suspicious (REQUIRED - MIN 2)\\
- {[}First potential issue/concern{]}\\
- {[}Second potential issue/concern{]}\\
\\
\#\# Arguments AGAINST Being Suspicious (REQUIRED - MIN 2)\\
- {[}First reason this might be normal{]}\\
- {[}Second reason this might be normal{]}\\
\\
\#\# Balanced Conclusion\\
{[}Weigh both sides and conclude whether this is suspicious or not{]}\\
\\
\#\# Root Cause vs Symptom\\
{[}If suspicious: Is this a root cause or symptom? If symptom, what's the root?{]}\\
\\
\#\# Fix Required\\
{[}If issue found: Specific code change needed. If not: 'No fix required'{]}"\\
~~\}\\
\}
\end{tcolorbox}
\caption{Graph Scanner prompt template (part 3 of 3).}
\label{fig:scanner-prompt-3}
\end{figure}

\subsection{Insight Rover Prompts}
\label{app:insight-rover}

The Insight Rover uses three distinct prompts for hypothesis generation, exploration planning, and node analysis within narrative context.

\begin{figure}[htbp]
\centering
\begin{tcolorbox}[
    colback=white,
    colframe=black,
    boxrule=0.5pt,
    left=5pt,right=5pt,top=5pt,bottom=5pt,
    width=\columnwidth,
    title=Insight Rover: Initial Hypothesis Generation Prompt,
    fonttitle=\bfseries,
    coltitle=black,
    colbacktitle=white
]
\footnotesize
Given a suspicious node in a hardware failure analysis, generate an initial hypothesis.\\
\\
Node: \{node.signal\_name\} at cycle \{node.cycle\}\\
Value: \{node.value\}\\
RTL Context: \{context.rtl\}\\
Spec Context: \{context.spec\}\\
\\
\textbf{\# Prior Analysis From GraphScanner}\\
\{prior\_analysis\_raw\}\\
\\
Generate a hypothesis about what might be wrong. Consider these possibilities:\\
- RTL bug (incorrect logic, missing conditions, wrong operators)\\
- Under-constrained inputs (missing assumptions in formal verification)\\
- Assertion/property issue (the checker itself might be wrong)\\
- Design intent mismatch (RTL correct but doesn't match specification)\\
\\
Generate a hypothesis in JSON format:\\
\{\\
~~~~"title": "Brief title for the hypothesis",\\
~~~~"hypothesis": "One-line hypothesis about what might be wrong (be specific about the type of issue)",\\
~~~~"initial\_insights": {[}"insight1", "insight2", ...{]}\\
\}
\end{tcolorbox}
\caption{Insight Rover prompt for generating initial hypotheses from suspicious nodes.}
\label{fig:rover-hypothesis-prompt}
\end{figure}

\begin{figure}[htbp]
\centering
\begin{tcolorbox}[
    colback=white,
    colframe=black,
    boxrule=0.5pt,
    left=5pt,right=5pt,top=5pt,bottom=5pt,
    width=\columnwidth,
    title=Insight Rover: Exploration Target Selection Prompt,
    fonttitle=\bfseries,
    coltitle=black,
    colbacktitle=white
]
\footnotesize
Given a narrative hypothesis and exploration frontier, select the most promising nodes to explore next.\\
\\
Narrative: \{narrative.hypothesis\}\\
Current confidence: \{narrative.confidence\_score:.2f\}\\
Events found: \{len(narrative.events)\}\\
\\
Exploration frontier:\\
- \{node\_id1\}: \{signal1\} = \{value1\}\\
- \{node\_id2\}: \{signal2\} = \{value2\}\\
- \{node\_id3\}: \{signal3\} = \{value3\}\\
{[}... up to 10 frontier nodes shown ...{]}\\
\\
Select up to 3 nodes that would best help validate or refute this hypothesis.\\
Return as JSON: \{"targets": {[}"node\_id1", "node\_id2", ...{]}\}
\end{tcolorbox}
\caption{Insight Rover prompt for selecting which frontier nodes to explore next.}
\label{fig:rover-exploration-prompt}
\end{figure}

\begin{figure}[htbp]
\centering
\begin{tcolorbox}[
    colback=white,
    colframe=black,
    boxrule=0.5pt,
    left=5pt,right=5pt,top=5pt,bottom=5pt,
    width=\columnwidth,
    title=Insight Rover: Node Analysis in Narrative Context,
    fonttitle=\bfseries,
    coltitle=black,
    colbacktitle=white
]
\footnotesize
Analyze this node in the context of the narrative hypothesis.\\
\\
Narrative: \{narrative.hypothesis\}\\
Current timeline:\\
C\{cycle1\}: \{signal1\} = \{value1\}\\
C\{cycle2\}: \{signal2\} = \{value2\}\\
{[}... last 5 events shown ...{]}\\
\\
Node to analyze: \{node.signal\_name\} at cycle \{node.cycle\}\\
Value: \{node.value\}\\
RTL Context: \{context.rtl\}\\
\\
\textbf{\# Prior Analysis From GraphScanner}\\
\{prior\_analysis\_raw\}\\
\\
\textbf{IMPORTANT:} When providing evidence, directly quote or reference specific facts from the RTL context above.\\
For example: "The RTL shows 'assign ready = valid \&\& !busy' which indicates..."\\
\\
Determine:\\
1. Is this node relevant to the narrative?\\
2. Does it support or contradict the hypothesis?\\
3. Is it part of the critical path?\\
4. Extract specific evidence from the provided context\\
\\
Return analysis as JSON with fields:\\
- is\_relevant: boolean\\
- is\_critical: boolean\\
- event\_description: string (if relevant)\\
- importance: float (0-1)\\
- evidence\_strength: float (0-1)\\
- evidence\_for: {[}list of SPECIFIC facts/quotes from the RTL context that support the hypothesis{]}\\
- evidence\_against: {[}list of SPECIFIC facts/quotes from the RTL context that contradict the hypothesis{]}\\
- new\_insights: {[}list of analytical insights based on the evidence{]}
\end{tcolorbox}
\caption{Insight Rover prompt for analyzing nodes within the context of a specific narrative hypothesis.}
\label{fig:rover-analysis-prompt}
\end{figure}

\subsection{Fix Generator Prompts}
\label{app:fix-generator}

The Fix Generator uses an ensemble approach with multiple prompting strategies. All strategies share a common base prompt and fix generation instructions.

\begin{figure}[htbp]
\centering
\begin{tcolorbox}[
    colback=white,
    colframe=black,
    boxrule=0.5pt,
    left=5pt,right=5pt,top=5pt,bottom=5pt,
    width=\columnwidth,
    title=Fix Generator: Core Instructions,
    fonttitle=\bfseries,
    coltitle=black,
    colbacktitle=white
]
\footnotesize
\textbf{IMPORTANT:} Please analyze this formal verification issue carefully and provide your response ONLY in the following JSON format:\\
\\
\{\\
~~~~"category": "RTL Bug" or "Under-Constraint" or "Over-Constraint",\\
~~~~"analysis": "Your detailed analysis of the issue including root cause and evidence",\\
~~~~"fixes": {[}\\
~~~~~~~~\{\\
~~~~~~~~~~~~"buggy\_code": "The EXACT problematic code snippet that needs to be fixed\\
~~~~~~~~~~~~~~~~~~~~~~~~~~~~(must be an exact substring from the original code)",\\
~~~~~~~~~~~~"code": "Your proposed fixed code that should replace the buggy code",\\
~~~~~~~~~~~~"description": "Explanation of what this fix does and why it addresses the root cause",\\
~~~~~~~~~~~~"confidence": 0.9,~~// A value between 0 and 1 indicating your confidence in this fix\\
~~~~~~~~~~~~"location": \{\\
~~~~~~~~~~~~~~~~"module": "Target module name",\\
~~~~~~~~~~~~~~~~"signal": "Target signal name",\\
~~~~~~~~~~~~~~~~"file": "Target file path",\\
~~~~~~~~~~~~~~~~"line": 42~~// Target line number where fix should be applied\\
~~~~~~~~~~~~\}\\
~~~~~~~~\},\\
~~~~~~~~// Please try to include at least 5 alternative fixes in RANKED ORDER\\
~~~~~~~~// The first fix should be your best solution (highest confidence)\\
~~~~~~~~// Each additional fix should be an alternative approach with decreasing confidence\\
~~~~~~~~\{\\
~~~~~~~~~~~~"buggy\_code": "The same EXACT problematic code that needs to be fixed",\\
~~~~~~~~~~~~"code": "An alternative fixed code that should replace the buggy code",\\
~~~~~~~~~~~~"description": "Explanation of this alternative approach",\\
~~~~~~~~~~~~"confidence": 0.8,~~// Lower confidence than your top solution\\
~~~~~~~~~~~~"location": \{ ... \}\\
~~~~~~~~\}\\
~~~~~~~~// Continue with more alternative approaches (3-5 total fixes)\\
~~~~{]}\\
\}
\end{tcolorbox}
\caption{Fix Generator JSON format and basic instructions (part 1 of 2).}
\label{fig:fix-prompt-1}
\end{figure}

\begin{figure}[htbp]
\centering
\begin{tcolorbox}[
    colback=white,
    colframe=black,
    boxrule=0.5pt,
    left=5pt,right=5pt,top=5pt,bottom=5pt,
    width=\columnwidth,
    title=Fix Generator: Critical Requirements,
    fonttitle=\bfseries,
    coltitle=black,
    colbacktitle=white
]
\footnotesize
\textbf{CRITICAL REQUIREMENTS:}\\
1. The "buggy\_code" field MUST contain EXACT code that exists in the RTL\\
2. The "code" field must contain ACTUAL CORRECTED RTL CODE - not empty, not placeholders\\
3. Both fields must contain valid Verilog/SystemVerilog syntax\\
4. Generate AT LEAST 3-5 different fixes if possible\\
5. Focus on FUNCTIONAL bugs that affect behavior - NOT style issues\\
6. Do NOT add testbench/verification signals (like \_assert)\\
7. Do NOT use angle brackets or placeholder text like "TODO", "TBD", etc.\\
8. Make sure the "buggy\_code" is an exact substring that exists in the original code\\
9. The "line" number should point to the approximate location of the buggy code\\
\\
\textbf{WHITESPACE HANDLING:}\\
- IMPORTANT: Pay careful attention to whitespace in the "buggy\_code" field\\
- Use SPACES instead of TABS in your code - avoid using \\t characters\\
- Try to match the whitespace pattern from the original RTL code\\
- When in doubt, use single spaces between tokens (e.g., "assign signal = value;")\\
- The validation will try to match with flexible whitespace, but exact matches work best\\
\\
\textbf{IMPORTANT NOTES:}\\
- Ensure your analysis thoroughly explains the root cause of the issue\\
- Provide multiple alternative fixes ranked from highest to lowest confidence\\
- Each fix should be a complete, working solution - no placeholders\\
- Consider different types of fixes:\\
~~* RTL bug (incorrect logic, missing conditions, wrong operators)\\
~~* Under-constrained inputs (missing assumptions in formal verification)\\
~~* Assertion/property issue (the checker itself might be wrong)\\
~~* Design intent mismatch (RTL correct but doesn't match specification)\\
\\
\textbf{STRICT SPAN RULES:}\\
- Use a SINGLE, MINIMAL SPAN for both fields (one assignment or contiguous lines only)\\
- Do NOT include case labels, begin/end, or asserts in buggy\_code or code\\
- We perform literal text replacement of buggy\_code with code; include only the exact text to replace\\
\\
Return either a JSON array of fixes or an object with a 'fixes' array.\\
Do NOT include any markdown code blocks or additional text outside the JSON.
\end{tcolorbox}
\caption{Fix Generator critical requirements and guidelines (part 2 of 2).}
\label{fig:fix-prompt-2}
\end{figure}

\begin{figure}[htbp]
\centering
\begin{tcolorbox}[
    colback=white,
    colframe=black,
    boxrule=0.5pt,
    left=5pt,right=5pt,top=5pt,bottom=5pt,
    width=\columnwidth,
    title=Fix Generator: Strategy-Specific Context Additions,
    fonttitle=\bfseries,
    coltitle=black,
    colbacktitle=white
]
\footnotesize
\textbf{Strategy: full\_context}\\
\#\# Key Insights from Analysis:\\
- \{insight1\}\\
- \{insight2\}\\
{[}... up to 10 insights ...{]}\\
\\
\#\# Most Suspicious Signals:\\
- Signal '\{signal\}' at cycle \{cycle\}: suspicion score \{score:.2f\}\\
~~Insights: \{insight1\}; \{insight2\}\\
{[}... up to 5 suspicious signals ...{]}\\
\\
\#\# Causal Analysis Narratives:\\
1. \{narrative1\}\\
2. \{narrative2\}\\
{[}... up to 3 narratives ...{]}\\
\\
\textbf{Strategy: suspicious\_focus}\\
\#\# CRITICAL: Focus on these suspicious signals:\\
- Signal '\{signal\}' (cycle \{cycle\}): HIGH SUSPICION (\{score:.2f\})\\
~~→ \{insight1\}\\
~~→ \{insight2\}\\
~~→ \{insight3\}\\
{[}... up to 7 suspicious signals with detailed insights ...{]}\\
Prioritize fixes for these highly suspicious signals!\\
\\
\textbf{Strategy: causal\_narratives\_focus}\\
\#\# Root Cause Narratives (FOCUS ON THESE):\\
\#\#\# Narrative 1:\\
\{full\_narrative1\}\\
\#\#\# Narrative 2:\\
\{full\_narrative2\}\\
{[}... up to 5 narratives ...{]}\\
\#\# Generate fixes that directly address the root causes identified in these narratives.\\
\\
\textbf{Strategy: minimal\_context}\\
\#\# Critical Issue:\\
ROOT CAUSE: \{root\_cause\_bug\}\\
Generate 3-5 surgical fixes for this specific issue.\\
\\
\textbf{Strategy: bugs\_and\_suggestions\_only}\\
\#\# Bugs and Fix Suggestions:\\
Signal '\{signal\}' bugs:\\
- \{bug1\}\\
- \{bug2\}\\
Suggested fixes:\\
- \{suggestion1\}\\
- \{suggestion2\}\\
{[}... up to 5 signals with bugs and suggestions ...{]}
\end{tcolorbox}
\caption{Examples of strategy-specific context additions for the Fix Generator ensemble approach.}
\label{fig:fix-strategies}
\end{figure}

\subsection{Narrative Ranking Prompts}
\label{app:ranking}

\fvdebug uses LLM-based ranking at two critical points: ranking competing narratives during exploration and evaluating hypothesis quality against ground truth during benchmarking.

\begin{figure}[htbp]
\centering
\begin{tcolorbox}[
    colback=white,
    colframe=black,
    boxrule=0.5pt,
    left=5pt,right=5pt,top=5pt,bottom=5pt,
    width=\columnwidth,
    title=Narrative Ranking: Intrinsic Quality Assessment,
    fonttitle=\bfseries,
    coltitle=black,
    colbacktitle=white
]
\footnotesize
You are an expert hardware verification engineer evaluating competing hypotheses for a formal verification failure.\\
\\
\textbf{PROBLEM DESCRIPTION:}\\
\{problem\_description\}\\
\\
Your task is to rank these hypotheses based on their intrinsic qualities that correlate with correctness, WITHOUT knowing the ground truth.\\
\\
\textbf{HYPOTHESES TO EVALUATE:}\\
---\\
HYPOTHESIS \#1 (ID: \{narrative\_id1\})\\
\{narrative\_text1\}\\
---\\
HYPOTHESIS \#2 (ID: \{narrative\_id2\})\\
\{narrative\_text2\}\\
{[}... additional hypotheses ...{]}\\
\\
\textbf{EVALUATION CRITERIA:}\\
\\
Score each hypothesis on these dimensions (0.0 to 1.0):\\
\\
1. \textbf{Sufficiency} (0.0-1.0): Does the hypothesis provide a complete explanation of the failure?\\
~~~- High (0.8-1.0): Fully explains the failure mechanism with clear causal chain\\
~~~- Medium (0.4-0.7): Partial explanation with some gaps\\
~~~- Low (0.0-0.3): Incomplete or superficial explanation\\
\\
2. \textbf{Evidence} (0.0-1.0): Quality and quantity of supporting evidence from the causal analysis\\
~~~- High (0.8-1.0): Strong evidence with specific signal values, RTL snippets, and clear causal links\\
~~~- Medium (0.4-0.7): Some evidence but lacks specificity or completeness\\
~~~- Low (0.0-0.3): Weak or contradictory evidence\\
\\
3. \textbf{Mechanistic Insight} (0.0-1.0): Clarity of the failure mechanism explanation\\
~~~- High (0.8-1.0): Clear explanation of HOW the bug manifests in hardware behavior\\
~~~- Medium (0.4-0.7): Some mechanistic understanding but unclear details\\
~~~- Low (0.0-0.3): Vague or incorrect understanding of hardware behavior\\
\\
4. \textbf{Actionability} (0.0-1.0): Does it provide clear guidance on what to fix?\\
~~~- High (0.8-1.0): Specific fix location and clear correction needed\\
~~~- Medium (0.4-0.7): General area identified but unclear exact fix\\
~~~- Low (0.0-0.3): No clear fix guidance\\
\\
5. \textbf{Coherence} (0.0-1.0): Internal consistency and logical flow\\
~~~- High (0.8-1.0): Logically consistent with no contradictions\\
~~~- Medium (0.4-0.7): Mostly consistent with minor issues\\
~~~- Low (0.0-0.3): Contains contradictions or illogical jumps\\
\\
\textbf{IMPORTANT CONSIDERATIONS:}\\
- Favor hypotheses that identify specific RTL bugs over vague constraint issues\\
- Value concrete evidence (specific signal values, code snippets) over speculation\\
- Prefer hypotheses with clear causal chains showing propagation of errors\\
- Penalize hypotheses that blame tools/extraction without strong justification\\
- Reward specificity about the exact issue and its location
\end{tcolorbox}
\caption{LLM prompt for ranking narratives based on intrinsic quality metrics without ground truth.}
\label{fig:ranking-prompt}
\end{figure}

\begin{figure}[htbp]
\centering
\begin{tcolorbox}[
    colback=white,
    colframe=black,
    boxrule=0.5pt,
    left=5pt,right=5pt,top=5pt,bottom=5pt,
    width=\columnwidth,
    title=Narrative Ranking: Output Format,
    fonttitle=\bfseries,
    coltitle=black,
    colbacktitle=white
]
\footnotesize
\textbf{OUTPUT FORMAT:}\\
Provide your evaluation as a JSON array where each element corresponds to a hypothesis:\\
{[}\\
~~~~\{\\
~~~~~~~~"hypothesis\_id": "ID of the hypothesis being evaluated",\\
~~~~~~~~"sufficiency": 0.85,\\
~~~~~~~~"evidence": 0.90,\\
~~~~~~~~"mechanistic\_insight": 0.80,\\
~~~~~~~~"actionability": 0.75,\\
~~~~~~~~"coherence": 0.95,\\
~~~~~~~~"overall\_score": 0.85,  // Average of the five scores\\
~~~~~~~~"reasoning": "Brief explanation of why these scores were assigned",\\
~~~~~~~~"rank\_suggestion": 1  // Your suggested rank (1 = best)\\
~~~~\},\\
~~~~\{\\
~~~~~~~~"hypothesis\_id": "ID of the second hypothesis",\\
~~~~~~~~"sufficiency": 0.60,\\
~~~~~~~~"evidence": 0.55,\\
~~~~~~~~"mechanistic\_insight": 0.50,\\
~~~~~~~~"actionability": 0.45,\\
~~~~~~~~"coherence": 0.70,\\
~~~~~~~~"overall\_score": 0.56,\\
~~~~~~~~"reasoning": "Explanation for this hypothesis",\\
~~~~~~~~"rank\_suggestion": 2\\
~~~~\}\\
{]}\\
\\
\textbf{CRITICAL:} Evaluate ALL hypotheses and return them in your suggested rank order (best first).
\end{tcolorbox}
\caption{JSON output format for narrative ranking results.}
\label{fig:ranking-output}
\end{figure}

% For benchmarking purposes, \fvdebug includes an evaluator that compares generated hypotheses against ground truth answers to measure accuracy.

\begin{figure}[htbp]
\centering
\begin{tcolorbox}[
    colback=white,
    colframe=black,
    boxrule=0.5pt,
    left=5pt,right=5pt,top=5pt,bottom=5pt,
    width=\columnwidth,
    title=Ground Truth Evaluation Prompt,
    fonttitle=\bfseries,
    coltitle=black,
    colbacktitle=white
]
\footnotesize
You are an expert hardware verification engineer evaluating hypothesis quality for debugging formal verification failures.\\
\\
\textbf{PROBLEM DESCRIPTION:}\\
\{problem\_description\}\\
\\
\textbf{GOLDEN ANSWER (Ground Truth):}\\
\{golden\_answer\}\\
\\
\textbf{HYPOTHESIS TO EVALUATE (Rank \#\{hypothesis\_rank\}):}\\
\{hypothesis\}\\
\\
\textbf{ADDITIONAL CONTEXT:}\\
- This hypothesis was ranked \#\{hypothesis\_rank\} in a list of hypotheses\\
- The golden answer above shows the correct root cause identification (could be RTL bug, constraint issue, property issue, etc.)\\
\\
\textbf{EVALUATION TASK:}\\
Score this hypothesis on the following dimensions (0.0 to 1.0):\\
\\
1. \textbf{Relevance} (0.0-1.0): Does it address the actual issue described in the golden answer?\\
2. \textbf{Preciseness} (0.0-1.0): Is it specific about the root cause? Does it correctly identify the exact issue?\\
3. \textbf{Causal\_Timeline} (0.0-1.0): Does it include a causal timeline or temporal analysis showing how the bug manifests over time? Higher scores for detailed cycle-by-cycle analysis.\\
4. \textbf{Correctness} (0.0-1.0): Does it correctly identify the root cause as shown in the golden answer?\\
\\
\textbf{SCORING GUIDELINES:}\\
- High relevance (0.8-1.0): Directly mentions the specific issue from the golden answer\\
- Medium relevance (0.4-0.7): Mentions related issues but not the specific root cause\\
- Low relevance (0.0-0.3): Vague or mentions unrelated issues\\
\\
- High preciseness (0.8-1.0): Specifically identifies the exact issue as in the golden answer\\
- Medium preciseness (0.4-0.7): Mentions the general area of the issue but lacks specifics\\
- Low preciseness (0.0-0.3): Vague statements without specific identification\\
\\
- High causal\_timeline (0.8-1.0): Includes detailed cycle-by-cycle timeline showing bug progression\\
- Medium causal\_timeline (0.4-0.7): Includes some temporal analysis or partial timeline\\
- Low causal\_timeline (0.0-0.3): No timeline or temporal analysis provided\\
\\
- High correctness (0.8-1.0): Correctly identifies the root cause as shown in golden answer\\
- Medium correctness (0.4-0.7): Partially correct but includes incorrect elements\\
- Low correctness (0.0-0.3): Incorrect or focuses on non-existent issues
\end{tcolorbox}
\caption{Evaluation prompt for comparing hypotheses against ground truth (part 1 of 2).}
\label{fig:eval-prompt-1}
\end{figure}

\begin{figure}[htbp]
\centering
\begin{tcolorbox}[
    colback=white,
    colframe=black,
    boxrule=0.5pt,
    left=5pt,right=5pt,top=5pt,bottom=5pt,
    width=\columnwidth,
    title=Ground Truth Evaluation Prompt (continued),
    fonttitle=\bfseries,
    coltitle=black,
    colbacktitle=white
]
\footnotesize
\textbf{IMPORTANT CONSIDERATIONS:}\\
- Score based on alignment with the golden answer, regardless of whether it's an RTL bug, constraint issue, or property issue\\
- Hypotheses that identify a different type of issue than the golden answer should receive lower scores\\
- Value specificity: hypotheses that identify the exact issue (e.g., specific condition, signal, or constraint) should score higher\\
- REWARD detailed causal timelines that show the bug's progression through cycles - this demonstrates thorough analysis\\
- Do NOT penalize verbosity if it provides valuable temporal analysis or causal chain information\\
\\
\textbf{OUTPUT FORMAT:}\\
Provide your evaluation in the following JSON format:\\
\{\\
~~~~"relevance": <float between 0.0 and 1.0>,\\
~~~~"preciseness": <float between 0.0 and 1.0>,\\
~~~~"causal\_timeline": <float between 0.0 and 1.0>,\\
~~~~"correctness": <float between 0.0 and 1.0>,\\
~~~~"overall": <float between 0.0 and 1.0 (average of the four scores)>,\\
~~~~"reasoning": "<brief explanation of scores>"\\
\}\\
\\
Analyze the hypothesis carefully and provide your JSON evaluation:
\end{tcolorbox}
\caption{Evaluation prompt for comparing hypotheses against ground truth (part 2 of 2).}
\label{fig:eval-prompt-2}
\end{figure}